\documentclass[12pt,a4paper]{article}
\usepackage[margin=2cm]{geometry}
\usepackage{authblk}

\usepackage[breaklinks]{hyperref}
\usepackage{booktabs}
\usepackage{txfonts}

\usepackage{todonotes,soul}
\usepackage{braket}

\usepackage{amsthm}
\usepackage{amsmath}
\usepackage{amssymb}
\usepackage{amsfonts}
\usepackage{mathbbol}
\usepackage{caption}
\usepackage{bm}
\usepackage{xcolor}
\usepackage{mathrsfs}

\newcommand{\be}{\begin{equation}}
\newcommand{\ee}{\end{equation}}
\newcommand{\bea}{\begin{eqnarray}}
\newcommand{\eea}{\end{eqnarray}}
\def\bml{\begin{subequations}}
\def\blea{\bml\begin{eqnarray}}
\def\eml{\end{subequations}}
\def\elea{\end{eqnarray}\eml}

\newtheorem{theorem}{Theorem}[section]
\newtheorem{corollary}[theorem]{Corollary}
\newtheorem{lemma}[theorem]{Lemma}
\newtheorem{proposition}[theorem]{Proposition}

\def\fmax{\phi_{\text{max}}}

\newcommand{\nord}[1]{{:}#1{:}}

\begin{document}

\title{A new derivation of singularity theorems with weakened energy hypotheses}

\author{Christopher J. Fewster\thanks{\tt chris.fewster@york.ac.uk} }
\author{Eleni-Alexandra Kontou\thanks{\tt eleni.kontou@york.ac.uk}}
\affil{Department of Mathematics, University of York, Heslington, York YO10 5DD, United Kingdom}
\date{\small\today}
\maketitle

\begin{abstract}
	The original singularity theorems of Penrose and Hawking were proved for matter obeying the Null Energy Condition or Strong Energy Condition respectively. Various authors have proved versions of these results under weakened hypotheses, by considering the Riccati inequality obtained from Raychaudhuri's equation. Here, we give a different derivation that avoids the Raychaudhuri equation but instead makes use of index form methods. We show how our results improve over existing methods and how they can be applied to hypotheses inspired by Quantum Energy Inequalities. In this last case, we make quantitative estimates of the initial conditions required for our singularity theorems to apply.
\end{abstract}

\section{Introduction}

A central question in gravitational physics is to determine conditions under which singularities arise either as the endpoint of gravitational collapse or at the origin of an expanding universe. Initial efforts to answer this question were restricted to spacetimes with high symmetry or simple matter models. A major breakthrough occurred with the proof of general singularity theorems by Penrose~\cite{Penrose:1964wq} and Hawking~\cite{Hawking:1966sx} in the mid-1960's. There, for the first time, it was proved that a singularity is inevitable if the spacetime and matter obey a series of general assumptions. It is striking that these revolutionary theorems were proven about 5 years before the identification of the first black hole candidate~\cite{WebsterMurdin:1972,Bolton:1972} and within months of the cosmic microwave background (CMB) discovery~\cite{PenziasWilson:1965} (the link with singularity theorems was discussed in~\cite{HawkingEllis:1968}).
Today, though observation now provides direct images of a black hole~\cite{EHTC:2019iv} and detailed CMB measurements~\cite{aghanim2018planck}, the question of whether singularities exist in our physical universe remains open. Of course it is impossible to observationally detect singularities and a full theoretical answer requires a consistent theory of quantum gravity. In the absence of such a theory, semiclassical gravity provides the most accurate current model of the universe. It is therefore important to seek improved versions of singularity theorems that take into account the properties of quantised matter in order to shed light in the necessary conditions that lead to singular spacetimes in a semiclassical approximation.

In the context of singularity theorems a spacetime is defined as singular if it contains at least one causal geodesic that is inextendible and incomplete to the future (or past). That is, it cannot be extended arbitrarily far to the future as an affinely parameterised geodesic. This situation need not be associated with a curvature singularity. The basic ingredients of the singularity theorems may be grouped under three headings~\cite{Senovilla:2018aav}: causality assumptions on the spacetime, energy conditions on its matter content, and an initial condition at some surface $S$. The causality assumptions are used to show that a future-geodesically complete spacetime must contain a future-complete geodesic that emerges normally from $S$ 
and has no focal points to $S$. The precise definition of a focal point will be recalled below in Sec.~\ref{sec:index}; for the moment it is enough to think of it as a point where nearby geodesics leaving $S$ normally focus (to a good approximation). On the other hand, the energy conditions and the initial condition are used to show that every future-complete geodesic emerging normally from $S$ must contain at least one focal point. It follows that no spacetime obeying the causality, energy condition and initial conditions can be future-geodesically complete.\footnote{There are also singularity theorems, such as the Hawking--Penrose theorem~\cite{HawkingPenrose:1970} that turn on the existence of a pair of conjugate points along a geodesic, but which we will not consider here.} 

This paper is concerned with the link between energy conditions and focal points. The standard approach, as presented in~\cite{Penrose:1964wq,Hawking:1966sx,HawkingEllis,Wald_gr,Senovilla:2018aav,Witten:2019qhl} for example, links the existence of focal points to the behaviour of certain geodesic congruences leaving $S$; a focal point exists if the expansion of the congruence diverges at finite affine parameter. The energy conditions are then used in conjunction with Raychaudhuri's equation to prove that such a divergence occurs, provided that the congurence is initially converging. Here, the energy conditions employed are typically the null energy condition (NEC), that 
$T_{\mu\nu}U^\mu U^\nu\ge 0$ for all null $U^\mu$ at all points in spacetime, or the strong energy condition (SEC), that $T_{\mu\nu}U^\mu U^\nu-T/(n-2)\ge 0$
for all timelike unit vectors $U^\mu$ at all points of spacetime, where $n$ is the spacetime dimension. 

There are good reasons to seek generalisations of these results with weaker conditions on the matter. At the microscale, for example, it is known that matter described by quantum fields cannot obey any pointwise energy conditions~\cite{Epstein:1965zza}. Meanwhile, on the macroscale, the SEC fails in the current era of our universe, due to the dominant effect of dark energy. Several authors have considered generalisations of the singularity theorems in this direction, starting from work of Tipler~\cite{Tipler:1978zz,Tipler:1978b} (see also \cite{ChiconeEhrlich:1980}) in which various averaged energy conditions are employed in place of the pointwise versions. Examples include~\cite{Borde:1987qr, Roman:1988vv, Wald:1991xn} and~\cite{Fewster:2010gm,BrownFewsterKontou:2018}. Most of these references involve the analysis of a Riccati inequality
\begin{equation}
\frac{D\theta}{dt}\le R_{\mu \nu} U^\mu U^\nu -\frac{\theta^2}{n-r}
\end{equation}
derived from the Raychaudhuri equation, using results such as those of~\cite{Galloway:1981}. 
Here $\theta$ is the expansion of the geodesic congruence with velocity field $U^\mu$ and 
$r=1$ (resp., $r=2$) for timelike (resp., null) geodesic congruences; our geometric conventions are stated at the end of this section. Similar techniques may be used to establish generalised versions of other results in mathematical relativity, e.g., the area theorem \cite{LeSourd_area:2018}.
 
In this paper, we will point out a more direct method for obtaining such results, which avoids the use of the Raychaudhuri equation and Riccati inequalities. Instead it is based on the study of the \emph{index form}, which arises as the the second variational derivative of the length functional about a geodesic. Our treatment has been influenced by O'Neill's discussion of the standard singularity theorems~\cite{ONeill}. Actually, index form methods were used by Chicone and Ehrlich~\cite{ChiconeEhrlich:1980} to prove the existence of conjugate points along complete geodesics, using the averaged null energy condition (ANEC) or the comparable condition for the SEC, but this seems to be the only occasion on which index forms have been used to establish singularity theorems under weakened conditions. Our treatment will make a much wider use of these techniques, and will prove results for local energy averages that are analogous to, but improve upon and are simpler to prove than, the results of~\cite{Fewster:2010gm,BrownFewsterKontou:2018}. 
 
There are two basic advantages to this approach. The first is that it works well with weakened integral energy conditions, more easily satisfied by classical and quantum fields. The second is that it gives an estimate of the proper time or affine parameter where the focal point is formed. That means we can estimate the location of the singularity, as well as the minimum contraction required to guarantee its existence. 

The averaged energy conditions we will study include some that are inspired by known quantum energy inequalities (QEIs) satisfied by quantum fields (which, as mentioned, cannot satisfy pointwise energy conditions in general). QEIs were first introduced by Ford~\cite{Ford:1978qya} and have since been established in a number of quantum field theory models in flat and curved spacetimes (see \cite{Fewster2017QEIs} for a recent review). Often such inequalities take the general form 
\be\label{eq:QEIgenform}
\int_\gamma \langle \rho \rangle_\omega f(\tau)^2 d\tau \geq -|||f|||^2 \,,
\ee
for $f$ a smooth compactly supported real-valued function. Here $\rho$ is the renormalized energy density or similar quantity along a timelike curve $\gamma$, $\omega$ is a Hadamard state, and $||| \cdot |||$ a Sobolev norm. 

It has been long known \cite{Fewster:1998pu} that the quantized minimally coupled scalar field admits a bound of this form with $|||f|||^2=(16\pi^2)^{-1} \|f''\|^2$ for averaging along a timelike geodesic in four-dimensional Minkowski space.  QEIs established for this field in curved spacetimes, e.g., Refs.~\cite{Fewster:2000,Fewster:2007rh} can be rewritten in this form. For example, Ref.~\cite{Kontou:2014tha} computed the QEI of~\cite{Fewster:2007rh} by perturbative methods,
writing the bound in terms of $L^2$-norms of derivatives of the averaging function and constants that depend on the upper bound of the curvature. The non-minimally coupled scalar field obeys (state-dependent) QEI bounds~\cite{Fewster:2007ec, Fewster:2018pey} that could potentially also be brought into the same form. Similar remarks apply to other fields for which QEIs have been found -- see~\cite{Fewster2017QEIs} for references.

This paper is organized as follows. In Sec.~\ref{sec:index} we present an overview of index form methods and derive the basic results leading to the proofs of the original Hawking and Penrose singularity theorems. In Sec.~\ref{sec:exponential} we show how the same method can be used to prove singularity theorems with weakened energy conditions. As an example we discuss the case of an exponential function for the timelike and null cases. In Sec.~\ref{sec:QEI} we present the main result of the paper which is the derivation of singularity theorems with energy conditions inspired by QEIs, cf.\ Eq~\eqref{eq:QEIgenform}. Approximations for the required initial conditions for geodesic incompleteness in different cases are derived. In Sec.~\ref{sec:applications} we apply the results of Sec.~\ref{sec:QEI} in the case of the classical non-minimally coupled Einstein-Klein-Gordon theory to show how our results can provide quantitative estimates. Finally we conclude in Sec.~\ref{sec:conclusions} with a summary and discussion of future work. 

\smallskip 
\noindent \emph{Conventions:} Unless otherwise stated, we consider a general spacetime dimension $n>2$ and adopt units in which $G=c=1$. We employ $[-,-,-]$ conventions in the Misner, Thorne and Wheeler classification \cite{MTW}. That is, the metric signature is $(+,-,-, \dots)$, the Riemann tensor is defined as $R^{\phantom{\lambda \eta\nu}\mu}_{\lambda \eta\nu}v^\nu=(\nabla_\lambda \nabla_\eta-\nabla_\eta \nabla_\lambda)v^\mu$, and the Einstein equation is $G_{\mu \nu}=-8\pi T_{\mu \nu}$. The d'Alembertian is written $\Box_g = g^{\mu\nu}\nabla_\mu\nabla_\nu$. 

\section{Index form methods and focal points}
\label{sec:index}

Here we present the basic theorems concerning the existence of focal points along timelike and null geodesics, in terms of index forms. At the end of each subsection we discuss how these theorems can be used to prove the Hawking and Penrose singularity theorems. Much of this section is closely related to chapters~10 and~14 of Ref.~\cite{ONeill}.
On a point of notation, differentiation with respect to a time parameter will be denoted with a dot, while differentiation with respect to an affine parameter on null geodesic is indicated with a prime.  

\subsection{Focal points along timelike geodesics}\label{sec:timelikefocal}

Let $(M,g_{\mu\nu})$ be a smooth Lorentzian spacetime, and let $S$ be a smooth spacelike hypersurface in $M$. If $\gamma:[0,\tau]\to M$ is a smooth timelike curve, its length (i.e., the total proper time along $\gamma$)
is
\begin{equation}
	L[\gamma] = \int_0^\tau |\dot{\gamma}(t)| \,dt,\qquad |V|:=\sqrt{g_{\mu\nu}V^\mu V^\nu}.
\end{equation}
Let $q\in I^+(S)$ be fixed; then $\gamma$ is a critical point of the length functional among unit-speed timelike curves joining $S$ to $q$ if
and only if it is an affine geodesic issuing normally from $S$. In more detail, $\gamma$ is
a geodesic of this type if and only if $dL[\gamma_s]/ds|_{s=0}=0$ for every 
smooth $1$-parameter family of curves $\gamma_s:[0,\tau]\to M$, $s\in (-\delta,\delta)$ obeying
\begin{equation}
	\gamma_0=\gamma, \qquad \gamma_s(0)\in S, \qquad \gamma_s(\tau)=q
\end{equation}
for all $s\in (-\delta,\delta)$. Introducing some terminology, the partial derivatives of $\gamma_s(t)$ (in an arbitrary system of coordinates) with respect to $t$ and $s$ determine the 
\emph{longitudinal} and \emph{transverse} vector fields $U^\mu = \partial \gamma_s(t)^\mu/\partial t$, $V^\mu = \partial \gamma_s(t)^\mu/\partial s$, obeying  
\begin{equation}\label{eq:UV}
\nabla_U V^\mu = \nabla_V U^\mu,
\end{equation}
an identity which holds for any smooth 1-parameter family of curves. 
In particular, the restrictions of $U^\mu$ and $V^\mu$ to $\gamma$ yield the velocity vector $U^\mu|_{\gamma(t)} = \dot{\gamma}^\mu(t)$ and \emph{variation vector field} $V^\mu|_{\gamma(t)} = d\gamma_s(t)^\mu/ds|_{s=0}$. 

The above discussion may be generalised to piecewise smooth curves, with the result that $\gamma$ is a critical point of $L$ among unit-speed piecewise smooth curves if and only if it is an unbroken timelike geodesic in its proper time parametrisation. The variation field arising from a piecewise smooth variation of $\gamma$ is continuous and piecewise smooth, while the velocity field may have discontinuities (see~\cite[\S 10.1]{BeemEhrlichEasley} for further background on piecewise smooth variations).  

Now suppose that $\gamma$ is an affine geodesic emanating normally from $S$. A point $p$ on $\gamma$ is a \emph{focal point to $S$ along $\gamma$} if there is a nontrivial variation of $\gamma$, among geodesics issuing normally from $S$, with a variation field that vanishes at $p$. As we consider variations among geodesics, the variation field obeys the equation of geodesic deviation, Eq.~\eqref{eq:Jacobi} below, i.e., it is a Jacobi field.

The existence of focal points along $\gamma$ is closely related to the question of whether  $\gamma$ is a local maximum of the length functional, among constant speed curves joining $S$ to $q$  (not necessarily geodesics). 
The analysis starts from a formula for the second derivative of the length functional for a variation $\gamma_s$ of $\gamma$ in which each $\gamma_s$ is a piecewise smooth constant speed curve\footnote{The speed $|\dot{\gamma}_s(t)|$ may vary with $s$, but not $t$.} joining $S$ to $q$:
\begin{equation}\label{eq:indexform}
	\left.\frac{d^2}{ds^2}L[\gamma_s]\right|_{s=0} = I[V]:=\int_0^\tau\left( \frac{DV^\mu}{dt}\frac{DV_\mu}{dt} - R_{\mu\nu\alpha\beta} U^\mu V^\nu V^\alpha U^\beta \right)\,dt + K_{\mu\nu}V^\mu V^\nu|_{\gamma(0)}.
\end{equation}
Here, the extrinsic curvature tensor of $S$ is defined by $K_{\mu\nu}=\nabla_\mu\xi_\nu$, where $\xi^\mu$ is the unit tangent field of the congruence of future-directed timelike geodesics emanating normally from $S$ (in particular $U^\mu|_\gamma=\xi^\mu|_\gamma$). The quantity $I[V]$ is called the \emph{index form}; it is usually presented in a polarised form as a bilinear form in two vector fields, but we will not need to do that here. 

Owing to the conditions placed
on $\gamma_s$, we have
\begin{equation}\label{eq:Vside}
	V^\mu U_\mu|_{\gamma(0)}= 0, \qquad V^\mu|_{\gamma(\tau)}=0, \qquad U_\mu\frac{DV^\mu}{dt}\equiv 0,
\end{equation}
where the first two conditions reflect the boundary conditions that $\gamma_s(0)\in S$, $\gamma_s(\tau)=q$, while the third is due to the restriction to constant speed curves. As $\gamma$ is an affine geodesic, these conditions imply that $U_\mu V^\mu\equiv 0$. 
Integrating by parts in~\eqref{eq:indexform}, it is not hard to see that the index form vanishes when $\gamma(\tau)$ is
a focal point to $S$ along $\gamma$ and $V^\mu$ is the corresponding Jacobi field, solving
\begin{equation}\label{eq:Jacobi}
\frac{D^2 V^\mu}{dt^2} + R^\mu_{\phantom{\mu}\nu\alpha\beta}U^\nu U^\alpha V^\beta=0\,.
\end{equation}
This is precisely the situation in which the second derivative test fails to determine whether $\gamma$ is a local maximum of $L$.  Table~\ref{tab:1034} summarises a more detailed relationship between the index form and focal points presented in Theorem 10.34 in \cite{ONeill}.
\begin{table}\begin{center}
		\begin{tabular}{ccc}
			$\not\exists$ focal point in $(0,\tau]$ & $\implies$ & $I[V]<0$ for all $V^\mu$ \\
			$\exists$ focal point in $(0,\tau)$ & $\implies$ & $I[V]>0$ for some $V^\mu$ \\
			only focal point in $(0,\tau]$ is $\tau$ & $\implies$ & $I[V]\le 0$ for all $V^\mu$, and $I[V]=0$ for some $V^\mu$
		\end{tabular}
	\end{center}\caption{Summary of Theorem~10.34 in \cite{ONeill}. Here `for some/all $V^\mu$' is to be interpreted as `for some/all piecewise smooth $V^\mu$ obeying the conditions~\eqref{eq:Vside}.}\label{tab:1034}
\end{table}
From these logical relationships, it may be seen that the first two implications may be replaced by `if and only if' statements. For if $I[V]<0$ for all $V^\mu$ then neither of the second or third conclusions holds, and therefore neither of the hypotheses on these lines can hold. Therefore there is no focal point in $(0,\tau]$ and so the first implication admits a converse. Similarly, if the second conclusion holds then neither of the first or third can and we deduce that there must be a focal point in $(0,\tau)$, so the implication on the second line also admits its converse. In particular, we have:
\begin{proposition}
	There exists a focal point $\gamma(r)$ to $S$ along $\gamma$ for $r\in (0,\tau]$ (resp., $r\in (0,\tau)$) if and only if $I[V]\ge 0$ (resp., $I[V]>0$) for some piecewise smooth $V^\mu$ obeying~\eqref{eq:Vside}. 
	Consequently: if $\gamma$ is length-maximising, then there is no focal point in $(0,\tau)$; if $\gamma$ is not length-maximising, then there is a focal point in $(0,\tau]$; if there is a focal point in $(0,\tau)$, then $\gamma$ is not length-maximising; if there is no focal point in $(0,\tau]$ then $\gamma$ is length-maximising.  
\end{proposition}

Next, suppose $v^\mu$ is a unit spacelike vector tangent to $S$ at $\gamma(0)$, extended along $\gamma$ by parallel transport. Then, any continuous piecewise smooth function $f$ obeying $f(0)=1$, $f(\tau)=0$ determines a continuous piecewise smooth variation field obeying~\eqref{eq:Vside} by
\begin{equation}
	V^\mu = f v^\mu.
\end{equation} 
Noting that $D V^\mu/dt = \dot{f}v^\mu$, which is spacelike, we have
\begin{equation}
	I[V] = \int_0^\tau\left( -\dot{f}^2 - f^2 R_{\mu\nu\alpha\beta} U^\mu v^\nu v^\alpha U^\beta \right)\,dt + K_{\mu\nu}v^\mu v^\nu|_{\gamma(0)}. 
\end{equation}
But now consider an orthonormal basis $e_\mu$ ($\mu=0,\ldots,n-1$) for the tangent space to $S$ at $\gamma(0)$, in which $e_0^\mu=U^\mu$, and apply the preceding argument to each element $e_i$ ($i=1,\ldots,n-1$), keeping the same scalar function $f$. Summing, we find
\begin{equation}
	\sum_{i=1}^{n-1} I[f e_i] =  -\int_0^\tau\left( (n-1)\dot{f}^2 + f^2 R_{\mu\nu} U^\mu U^\nu  \right)\,dt - K|_{\gamma(0)},
\end{equation}
where we have used the fact that $U^\mu U^\nu K_{\mu\nu}=0$, so that the extrinsic curvature $K=g^{\mu\nu}K_{\mu\nu}=-\sum_i K(e_i,e_i)$. If there is no focal point in $(0,\tau]$ (resp., $(0,\tau)$) then each term on the left-hand side is negative (resp., nonpositive) for all $f$ obeying the boundary conditions $f(0)=1$, $f(\tau)=0$. Conversely, if the right-hand side is nonnegative (resp., positive) for some such $f$, then the same must be true of at least one of the terms on the left, and it follows that there is
a focal point in $(0,\tau]$ (resp., $(0,\tau)$). In other words, we have sufficient conditions for the presence of a focal point in $(0,\tau)$ or $(0,\tau]$ (and consequently some necessary conditions for their absence).  

\begin{proposition}
	\label{prop:timelike}
	Let $\gamma:[0,\tau]\to M$ be a unit-speed timelike geodesic emanating normally from a smooth spacelike hypersurface $S$. 
	If there exists a continuous, piecewise smooth $f$ on $[0,\tau]$ obeying $f(0)=1$, $f(\tau)=0$ and
	\begin{equation}\label{eq:proptimelike}
		\int_0^\tau\left( (n-1)\dot{f}^2 + f^2 R_{\mu\nu} U^\mu U^\nu  \right)\,dt\le -K|_{\gamma(0)},
	\end{equation}
	then there is a focal point to $S$ along $\gamma$. If the inequality~\eqref{eq:proptimelike} holds with a strict inequality then the focal point lies before $\gamma(\tau)$. 
\end{proposition}
This result may be used to deduce the existence of focal points, without using the Raychaudhuri equation \cite[Prop.~10.37]{ONeill}. Recall that the timelike convergence condition asserts that $R_{\mu\nu} U^\mu U^\nu\le 0$ for all timelike $U^\mu$, which is equivalent to the SEC for solutions to the Einstein equations.
\begin{corollary}\label{cor:timelikefocal}
	If  $K|_{\gamma(0)}<0$, $\tau\ge (n-1)/|K|_{\gamma(0)}|$, and the Ricci tensor obeys the timelike convergence criterion, then there is a focal point to $S$ along $\gamma$, i.e., by proper time $\tau$ at the latest. If $\tau> (n-1)/|K|_{\gamma(0)}|$ the focal point occurs before time $\tau$.
\end{corollary}
\begin{proof}
	Apply Prop.~\ref{prop:timelike} using the function $f(t)=1-t/\tau$, noting that the left-hand side of~\eqref{eq:proptimelike} is less than or equal to $(n-1)/\tau$ which is less than or equal to  $-K|_{\gamma(0)}$ by assumption. This gives the first stated result; the second is a trivial modification.
\end{proof}

At this point it is helpful to explain the relationship between the index form method and the traditional approaches based on the Raychaudhuri equation. One way to determine whether the condition given in Prop.~\ref{prop:timelike} holds is to solve the variational problem of minimising the right-hand side of 
Eq.~\eqref{eq:proptimelike}, treated as a functional $J[f]$, over smooth $f$ obeying $f(0)=1$, $f(\tau)=0$. 
Considering variations $f+\epsilon g$ where $g$ is smooth and obeys $g(0)=g(\tau)=0$, one finds easily that
\begin{equation}
J[f+\epsilon g] = J[f]+ 2\epsilon \int_0^\tau\left(- (n-1)\ddot{f} + f \rho  \right)g \,dt
+ \epsilon^2 J[g]\,,
\end{equation}
where we now write $\rho(t)=R_{\mu\nu} U^\mu U^\nu|_{\gamma(t)}$. 
There is at most one stationary point, namely the solution to 
\begin{equation}\label{eq:EulerLagrange}
- (n-1)\ddot{f} + f \rho = 0, \qquad f(0)=1,~f(\tau)=0\,,
\end{equation} 
if it exists (which it does unless there is a solution to the same equation with $f(0)=f(\tau)=0$).
For this solution, one finds $J[f] = (n-1)\dot{f}(0)$ using an integration by parts.\footnote{This stationary point is the global minimum of $J$ provided that $J[g]\ge 0$ for all smooth $g$ obeying Dirichlet boundary conditions, which can be analysed as an eigenvalue problem. However we will not need to do this.}
Therefore a sufficient condition for~\eqref{eq:proptimelike} to hold is that the solution to~\eqref{eq:EulerLagrange} has $\dot{f}(0)\le -K|_{\gamma(0)}/(n-1)$. 

Assuming that $f$ is nonvanishing in $(0,\tau)$, we may now set $\theta=(n-1)\dot{f}/f$ and note that the Euler--Lagrange equation~\eqref{eq:EulerLagrange} may be rewritten as the Riccati equation
\begin{equation}\label{eq:Riccati}
\dot{\theta} = -\frac{\theta^2}{n-1} +\rho\,, \qquad \theta(0) = (n-1)\dot{f}(0)\,,
\end{equation}
with $\theta\to-\infty$ as $t\to \tau^-$. (If $f$ has an interior zero then we repeat the reasoning on the interval between $t=0$ and the first zero.) We see that a sufficient condition for the existence of a focal point along $\gamma$ is that the above Riccati equation fails to have a solution beyond $[0,\tau]$ for initial data with $\theta(0)\le -K|_{\gamma(0)}$.

By contrast, now consider the (irrotational) congruence of unit speed timelike geodesics emanating normally from the $S$ with velocity field $U^\mu$. The Raychaudhuri equation for the expansion $\theta = \nabla_\mu U^\mu$ gives
\begin{equation} 
\frac{D\theta}{dt} =R_{\mu \nu} U^\mu U^\nu-2\sigma^2 -\frac{\theta^2}{n-1} ,\qquad \theta|_{\gamma(0)} = K|_{\gamma(0)}\,,
\end{equation}
along any geodesic in the congruence, where $\sigma$ is the shear scalar, and so $\theta$ obeys the differential inequality  
\begin{equation} \label{eq:Ricc}
\frac{D\theta}{dt}\le R_{\mu \nu} U^\mu U^\nu -\frac{\theta^2}{n-1}  ,\qquad \theta|_{\gamma(0)} = K|_{\gamma(0)}\,.
\end{equation}  
Therefore the Raychaudhuri equation leads to a similar analysis to that arising from the index form methods,
though with the slight complication of working with a differential inequality rather than a differential equation. More significantly, in the index form approach one need not pass to differential equations at all, but instead try to satisfy~\eqref{eq:proptimelike} by judicious choice of trial functions $f$. This is exactly what was done in the proof of Cor.~\ref{cor:timelikefocal} and will be our approach in the rest of this paper.

To complete this section, we now recall one of the simplest links between the presence of focal points and timelike geodesic incompleteness. The following result draws on~\cite{Galloway:1986} and~\cite[Thm 14.55A]{ONeill}. 
\begin{proposition}
	\label{prop:timelikeincom}
Let $S$ be a smooth spacelike Cauchy surface in $M$. (a) If $S$ is compact and every future-complete timelike geodesic emanating normally from $S$ contains a focal point then $M$ is future timelike geodesically incomplete. (b) If every future-directed timelike geodesic emanating normally from $S$ 
without focal points has length (strictly) less than $\tau_*$ then every future-directed timelike curve emanating from $S$ has length (strictly) less than $\tau_*$; consequently $M$ is future timelike geodesically incomplete. 
\end{proposition}
\begin{proof}
	(a) Assume that $M$ is future timelike geodesic complete. As discussed in the proof of~\cite[Thm 5.1]{Fewster:2010gm}, based on results of~\cite{Galloway:1986}, the hypotheses imply the existence of an $S$-ray; namely, a future inextendible unit-speed timelike geodesic $\gamma$ emanating (necessarily normally) from $S$ and which is length-maximising between $S$ and each of its points. Accordingly $\gamma$ contains no focal points to $S$, but as it is future-complete by assumption we obtain a contradiction. 
	
	(b) By~\cite[Lem.~14.29 \& Thm~14.44]{ONeill}, every $q\in D^+(S)\setminus S$ is connected to $S$ by a timelike length-maximising geodesic, which necessarily emanates normally from $S$ and has no focal points before $q$. By hypothesis, this geodesic has length (strictly) less than $\tau_*$.
	
	Now consider any unit-speed future-directed timelike curve $\gamma:[0,\tau]\to M$ with $\gamma(0)\in S$, assuming for a contradiction that $\tau> \tau_*$ (or $\tau\ge\tau_*$). 
	But $\gamma(\tau)$ is also joined to $S$ by a length-maximising geodesic of length $\tau'$ (strictly) less than $\tau_*$, so we deduce $\tau_*\ge\tau'\ge \tau>\tau_*$ (or $\tau_*>\tau'\ge \tau\ge\tau_*$) thus
	obtaining a contradiction. In particular, every inextendible future-directed timelike geodesic emanating from $S$ has finite length and is therefore incomplete.
\end{proof}
Combining Corollary~\ref{cor:timelikefocal} and Prop.~\ref{prop:timelikeincom} yields one of Hawking's singularity theorems~\cite{Hawking:1966sx}.
\begin{corollary}
	\label{cor:timelikeincom}
	If $\sup_S K<0$ on $S$ (in particular, if $S$ is compact and $K<0$) and the timelike convergence condition holds, then $M$ is future-timelike geodesically incomplete.
\end{corollary} 

The central idea of this paper is that results similar to  Cor.~\ref{cor:timelikefocal}, Prop.~\ref{prop:timelikeincom} and Cor.~\ref{cor:timelikeincom} may be proved under weaker assumptions on the Ricci tensor, by replacing the linear function $f$ in the proof of Cor.~\ref{cor:timelikefocal} by a suitable alternative. This will be described in Sections~\ref{sec:exponential} and~\ref{sec:QEI} below.
Before that, we describe how the above theory can be adapted to null geodesics, again following~\cite{ONeill}.

\subsection{Focal points along null geodesics}
\label{sub:focalnull}

Let $\gamma : [0,\ell] \to M$ be a piecewise smooth curve. Then the integral
\be\label{eq:energyint}
E[\gamma]=\frac{1}{2} \int_0^\ell g(\gamma'(\lambda),\gamma'(\lambda)) d\lambda \,,
\ee
is called the energy or action integral, and is the natural quantity to consider in the variational theory of null geodesics because, unlike the situation for $L$, $E[\gamma_s]$ varies smoothly in $s$ for any piecewise smooth variation of $\gamma$, regardless of the causal nature of $\gamma_s$. Let $P$ be a smooth semi-Riemannian submanifold of $M$, $q \in M \setminus P$ and $\Omega(P,q)$ the manifold of all piecewise smooth curve segments from $P$ to $q$. Then the critical points of $E$ among curves in $\Omega(P,q)$ are geodesics emanating normally from $P$, and the second derivative of $E$ in a smooth variation $\gamma_s$ of such a geodesic $\gamma=\gamma_0$ is 
\cite[Prop.~10.39]{ONeill} 
\be\label{eqn:Esecondder}
\frac{\partial^2 E[\gamma_s]}{\partial s^2} \bigg|_{s=0}=\int_0^\ell\left[(\nabla_U V_\mu)(\nabla_U V^\mu) -R_{\mu\nu \alpha \beta}U^\mu V^\nu V^\alpha U^\beta \right] d\lambda+(U_\mu \nabla_V V^\mu)\bigg|^{\ell}_{0} \,,
\ee
where $U^\mu=\gamma'(\lambda)$ and $V^\mu|_{\gamma(\lambda)} = d\gamma_s(\lambda)^\mu/ds|_{s=0}$, and
we assume that $\gamma$ is affinely parametrised.
The right-hand side of~\eqref{eqn:Esecondder} is, by definition, the Hessian $\mathcal{H}[V]$ of $E$ at critical points; it may also be written  
\be
\mathcal{H}[V]=\int_0^\ell \left(  \frac{D V^\mu}{d\lambda} \frac{D V_\mu}{d\lambda} -R_{\mu\nu \alpha \beta}U^\mu V^\nu V^\alpha U^\beta  \right)d\lambda-U_\mu \mathrm{I\!I}^\mu (V,V) \bigg|_{\gamma(0)} \,,
\ee
where $\mathrm{I\!I}$ is the shape tensor or second fundamental form, defined so that
$\mathrm{I\!I}^\mu (V,W)$ is the projection of $\nabla_V W^\mu$ onto the subspace of vectors normal to $P$, for vector fields $V,W$ tangential to $P$. The following result is proved as Proposition 10.41 of Ref.~\cite{ONeill}.
\begin{proposition}\label{prop:10.41}
Let $P$ be a spacelike submanifold of $M$. If there are no focal points of $P$ along a normal null geodesic $\gamma \in \Omega(P,q)$, then $\mathcal{H}[V]$ is positive semidefinite when restricted to
piecewise smooth $V^\mu$ obeying $U_\mu V^\mu\equiv 0$. If $\mathcal{H}[V]=0$ for some such $V^\mu$, then $V^\mu$ is tangent to $\gamma$.
\end{proposition}  

Now let $\gamma:[0,\ell]\to M$ be a null geodesic affinely parametrized by $\lambda$ and $P$ a spacelike $(n-2)$-dimensional submanifold of $M$. Let $e_i$ ($i=1, \dots ,n-2)$ be an orthonormal basis at $T_{\gamma(0)} (P)$. We parallel transport these vectors along $\gamma$ to get $E_i$ $(i=1, \dots ,n-2)$. Let $f$ be a smooth function with $f(0)=1$ and $f(\ell)=0$. Then
\be
H[f E_i]=\int_0^{\ell} \left( -f'^2 -f^2 R_{\mu\nu \alpha \beta}U^\mu E_i^\nu E_i^\alpha U^\beta  \right)d\lambda-U_\mu \mathrm{I\!I}^\mu (E_i,E_i) \bigg|_{\gamma(0)}\,.
\ee
Now sum over all $i=1, \dots, n-2$, noting that $g^{\nu\alpha} = U^{(\nu} W^{\alpha)}-\sum_{i=1}^{n-2}E_i^\nu E_i^\alpha$
for a suitably chosen null vector $W$, to obtain
\be\label{eq:prenull}
\sum_{i=1}^{n-2} \mathcal{H}[f E_i]=-\int_0^{\ell} \left( (n-2)f'^2 -f^2 R_{\mu\nu }U^\mu  U^\nu \right)d\lambda-(n-2)U_\mu H^\mu |_{\gamma(0)} \,,
\ee
where 
\begin{equation}
H^\mu=\frac{1}{n-2}\sum_{i=1}^{n-2}\mathrm{I\!I}^\mu (E_i,E_i)
\end{equation} 
is the mean normal curvature vector field of $P$. 

This calculation may be used in conjunction with Prop.~\ref{prop:10.41} to give a sufficient condition for the existence of a focal point along $\gamma$, just as in the derivation of Prop.~\ref{prop:timelike}. As there is no unique natural parametrisation of a null geodesic, it is convenient to state the result in an invariant form, regarding $\gamma$ as an unparametrised $1$-dimensional submanifold of $M$. 
The notation $d\gamma^\mu$ denotes the line element $1$-form on $\gamma$,
giving the tangent vector $d\gamma^\mu/d\lambda$ with respect to any coordinate $\lambda$ on $\gamma$, while $d\gamma^\mu_+$ is the pseudo-$1$-form which is
equal to $d\gamma^\mu$ with respect to coordinates parametrising $\gamma$ as a future-directed curve. Expressions such as the right-hand side of~\eqref{eq:prenull} may be written in invariant form by regarding $f(\lambda)$ as the coordinate expression of a density $f$ of weight $-\tfrac{1}{2}$ on $\gamma$. Note that the combination $f^2 d\gamma_+$ then defines a vector field along $\gamma$, given in coordinates by $f(\lambda)^2 d\gamma^\mu/d\lambda$, provided that $d\gamma^\mu/d\lambda$ is future-pointing. It should be borne in mind that, while a density does not have invariant values at individual points, the sign of the density is invariantly defined; likewise, it makes sense to say that the density vanishes or is nonvanishing at a given point. In this notation, the result we have proved may be stated as follows. 
\begin{proposition}
	\label{prop:null}
	Let $P$ be a spacelike submanifold of $M$ of co-dimension $2$ and let $\gamma$ be a null geodesic joining $p\in P$ to $q\in J^+(P)$. If there exists a smooth $(-\tfrac{1}{2})$-density $f$ on $\gamma$ which is nonvanishing at $p$ but vanishes at $q$ and so that
	\be
	\label{eqn:nullgen}
	\int_\gamma \left( (n-2)(\nabla_{d\gamma}f)^2 +f^2 \mathrm{Ric}(d\gamma,d\gamma) \right)   \leq -(n-2) g(f^2 d\gamma_+, H) |_{p} \,,
	\ee
	then there is a focal point to $P$ along $\gamma$; if the inequality holds strictly, then the focal point is located before $q$. 
\end{proposition}
Note that $|\nabla_{d\gamma} f|$ is a $\tfrac{1}{2}$-density on $\gamma$, while $\mathrm{Ric}(d\gamma,d\gamma)$ is a $2$-density. Accordingly, each term in the integrand of~\eqref{eqn:nullgen} is a density; similarly, $f^2d\gamma^\mu_+$ is a vector field along $\gamma$.
Written more explicitly, if $\gamma$ is parametrised by a coordinate $\lambda\in [0,\ell]$, inequality~\eqref{eqn:nullgen} is   
\be 
\int_0^\ell  \left( (n-2)f'(\lambda)^2 +f(\lambda)^2 R_{\mu\nu}U^\mu U^\nu \right) d\lambda \leq -(n-2)f(0)^2 U_\mu H^\mu|_{p} \,,
\ee
assuming $U^\mu=d\gamma^\mu/d\lambda$ is future-directed.

Proposition~\ref{prop:null} may be used to prove the following Corollary (cf.~\cite[Prop.~10.43]{ONeill}). Recall that $P$ is said to be \emph{future-converging} if $H^\mu$ is past-pointing timelike everywhere on $P$.
In this situation we may write $H^\mu=H \hat{H}^\mu$ where $H<0$ and $\hat{H}^\mu$ is a future-pointing timelike unit vector. For any future-pointing timelike unit vector $V^\mu$ at $p$, we also write $L_V(\gamma)$ for the length of $\gamma$ with respect to an affine parameter in which $V_\mu d\gamma^\mu/d\lambda =1$ at $p$. We refer to $L_V(\gamma)$ as the \emph{$V$-length} of $\gamma$.
\begin{corollary}
\label{cor:penrfocal}
With $P$ and $\gamma$ as in Prop.~\ref{prop:null}, suppose additionally that $P$ is future-converging and the null convergence condition  $\mathrm{Ric}(d\gamma,d\gamma) \leq 0$ holds everywhere along $\gamma$. If 
$L_{\hat{H}}(\gamma)  \geq 1/|H|$ then there is a focal point to $P$ along $\gamma$.  
\end{corollary}
\begin{proof}
	Choose an affine coordinate $\lambda$ on $\gamma$, so that $p=\gamma(0)$ 
	and $\hat{H}_\mu d\gamma^\mu/d\lambda =1$. Then 
	$q=\gamma(\ell)$, with $\ell=L_{\hat{H}}(\gamma)$. In these coordinates define $f(\lambda)=1-\lambda/\ell$; then the right-hand side of~\eqref{eqn:nullgen} is $-(n-2)H$, while the left-hand side is less than or equal to $(n-2)/\ell$,
	and the result follows by Proposition \ref{prop:null}. 
\end{proof}
As in Sec.~\ref{sec:timelikefocal}, inequality~\eqref{eqn:nullgen} may be connected to a Riccati equation related to the Raychaudhuri equation for a null geodesic congruence.

Now we can connect the formation of focal points with future null geodesic incompleteness in the following way, drawing on the formulations in~\cite{Fewster:2010gm,ONeill}.
\begin{proposition}
	\label{prop:nullincompl}
	Suppose that: (i) $M$ is globally hyperbolic with non-compact Cauchy hypersurfaces; (ii) $P$ is a compact achronal smooth spacelike submanifold of $M$ of co-dimension $2$; and
	(iii) every future-complete null geodesic emanating normally from $P$ contains a focal point to $P$. Then $M$ is future null geodesically incomplete.  If (iii) is replaced by: (iii)${}'$ $P$ is future converging and every future-directed null geodesic emanating normally from $P$ with $\hat{H}$-length at least $\ell$ contains a focal point to $P$, then there is an inextendible null geodesic emanating normally from $P$ with $\hat{H}$-length less than $\ell$. 
\end{proposition}
\begin{proof}
	As described in the proofs of Theorem~5.2 in~\cite{Fewster:2010gm} (see also the comparable part of \cite[Thm~14.61]{ONeill}) conditions~(i) and~(ii) imply that $E^+(P):=J^+(P)\setminus I^+(P)$ is equal to the boundary $\partial J^+(P)$ and is noncompact and closed. These properties were used to show that there is an inextendible affinely parametrised null geodesic $\gamma:[0,a)\to M$ issuing normally from $P$ and contained entirely in $\partial J^+(P)$ for some $a\in (0,\infty]$.
	Furthermore, $\gamma$ can contain no focal points to $P$, because the portion of $\gamma$ beyond any focal point would lie in $I^+(P)$~\cite[Prop.~10.48]{ONeill}, and hence outside $E^+(P)$. Assumption (iii) then entails that $\gamma$ is not future-complete and the result is proved. Alternatively, (iii)${}'$ implies immediately that $L_{\hat{H}}(\gamma)<\ell$.
\end{proof} 
The last two results combine to yield the Penrose singularity theorem~\cite{Penrose:1964wq}.
\begin{corollary}
	If, in addition to the assumptions (i) and (ii) of Prop.~\ref{prop:nullincompl}, $P$ is future-converging and the null convergence condition holds, then $M$ is future null geodesically incomplete.
\end{corollary}
\begin{proof}
	Corollary~\ref{cor:penrfocal} implies assumption (iii) of Prop.~\ref{prop:nullincompl}.
\end{proof}

\section{Exponential damping}
\label{sec:exponential}

The main goal of this paper is to show how the index form methods described in Propositions \ref{prop:timelike} and \ref{prop:null} can be used to prove singularity theorems with weaker energy conditions than the SEC or NEC, using much simpler arguments than those used in existing literature. Instead of controlling the (non)existence of solutions to the Raychaudhuri equation, the main method used here is to replace the linear functions used in the proof of Corollaries~\ref{cor:timelikefocal} and~\ref{cor:penrfocal} by functions that are adapted to the weakened energy conditions under consideration. Our first examples concern exponentially damped half-line averages of the timelike and null convergence conditions (corresponding to the SEC and NEC respectively), providing similar overall results to those derived in~\cite{Fewster:2010gm} but with much greater ease.

Starting with timelike geodesics, the following result generalises Corollaries~\ref{cor:timelikefocal} and~\ref{cor:timelikeincom} by weakening the timelike convergence condition.
It may be compared with Lem 3.1 in~\cite{Fewster:2010gm} combined with Theorem~5.1 of the same reference (modified as described in remark (1) following its proof). Our argument here represents a considerable simplification. 
\begin{theorem}
	Suppose $\gamma:[0,\infty)\to M$ is a future-directed unit-speed timelike geodesic emanating normally from a smooth spacelike hypersurface $S$. If the inequality
	\begin{equation}\label{eq:expsingbd}
	\frac{c}{2} + \liminf_{\tau\to+\infty}  \int_0^\tau (1-t/\tau)^2 e^{-2ct/(n-1)} R_{\mu\nu} U^\mu U^\nu \,dt < -K|_{\gamma(0)}
	\end{equation}
	holds for some $c\ge 0$, then there is a focal point to $S$ along $\gamma$. 
	
	If $S$ is additionally a compact Cauchy surface and~\eqref{eq:expsingbd} holds along
	every future-complete timelike unit-speed geodesic emanating normally from $S$ (the value of $c$ may vary) then $M$ is future timelike geodesically incomplete.
\end{theorem}
\begin{proof}
Define $F(\tau)$ to be the integral in Eq.~\eqref{eq:expsingbd}. Then there exists $\epsilon >0$ and a sequence $\tau_n \to \infty$ for which
\begin{equation}
	\label{eqn:taun}
	F(\tau_n)< - \frac{c}{2}- K|_{\gamma(0)}-\epsilon \,.
\end{equation}
Now define
\begin{equation}
	f_\tau(t)=(1-t/\tau)e^{-ct/(n-1)}
\end{equation} 
and note that
\begin{equation}
	\lim_{\tau\to +\infty} \int_0^\tau  (n-1)\dot{f}_\tau(t)^2 \,dt= \frac{c}{2} \,.
\end{equation}
Then from Eq.~\eqref{eqn:taun} for $\tau=\tau_n$ with $n$ sufficiently large we have
\begin{equation}\label{eq:ineq}
	\int_0^\tau  \left((n-1)\dot{f}_\tau(t)^2  + f_\tau(t)^2 R_{\mu\nu} U^\mu U^\nu \right)\,dt <  \frac{c}{2} + \epsilon  -\frac{c}{2} -K|_{\gamma(0)} - \epsilon = -K|_{\gamma(0)}
\end{equation}
and therefore there is a focal point before proper time $\tau$ by Prop.~\ref{prop:timelike}. The second part of the Theorem follows immediately from Prop.~\ref{prop:timelikeincom}(a). 
\end{proof}

The analogue for null geodesics is:
\begin{theorem}
	Assume that $P$ is a future convergent smooth spacelike codimension-$2$ submanifold of $M$. Suppose $\gamma:[0,\infty)\to M$ is an affinely parametrised null geodesic issuing normally from $P$, with $\hat{H}_\mu d\gamma^\mu/d\lambda=1$ for $\lambda=0$.  
	If the inequality
	\begin{equation} 
	\label{eq:nullexpsingbd}
	\frac{c}{2}+\liminf_{\ell \to +\infty} \int_0^\ell (1-\lambda/\ell)^2 e^{-2c\lambda/(n-2)} R_{\mu \nu} U^\mu U^\nu d\lambda< -(n-2) H 
	\end{equation}
	holds for some $c\ge 0$, then there is a focal point to $S$ along $\gamma$. 
	
	If, additionally, $M$ has a non-compact Cauchy surface, $P$ is compact, and~\eqref{eq:nullexpsingbd} holds along every future-complete null geodesic emanating normally from $P$, parametrised as described above (the value of $c$ may vary) then $M$ is future null geodesically incomplete.
\end{theorem}
The proof is exactly analogous to the timelike case, but making use of the function
$f_{\ell}(\lambda)=\left( 1-\lambda/\ell\right) e^{-c\lambda/(n-2)}$ and
Props.~\ref{prop:null} and~\ref{prop:nullincompl} in place of Props.~\ref{prop:timelike} and~\ref{prop:timelikeincom}(a).  

\section{QEI inspired hypotheses}
\label{sec:QEI}

In this section we replace the classical energy conditions with QEI-inspired hypotheses that can be obeyed by quantum fields. The main goal is to specify the required initial contraction that leads to geodesic incompleteness. 

We will study two scenarios: in the first we suppose that the timelike or null convergence condition is satisfied for a small segment of the geodesic; in the second, we impose conditions on the timelike- or null-contracted Ricci tensor at small negative values of proper time or the affine parameter respectively. Its worth noting that Ref.~\cite{Fewster:2010gm} also utilises the same information as in our second scenario to prove singularity theorems. 

Our method makes use of trial functions related to incomplete Beta functions, which we now define. For each $m\in \mathbb{N}$, let $p_m$ be the unique polynomial of degree $2m-1$ with the properties $p_m(0)=0$, $p_m(1)=0$, $p_m^{(k)}(0)=p_m^{(k)}(1)=0$ for $1\le k\le m-1$,
given explicitly by
\begin{equation}
	p_m(x) =\frac{1}{B(m,m)}\int_0^x y^{m-1}(1-y)^{m-1}\,dy\,,
\end{equation}
where $B$ is the Beta function. This polynomial is the regularised incomplete Beta function, denoted $p_m(x)=I(m,m;x)$ (the notation $I_x(m,m)$ is more common but less convenient for us)~\cite[\S 8.17]{NIST:DLMF}. In Appendix~\ref{appx:Sobolev} it is shown how the squares of the $L^2$ norms of $p_m$, $p_m'$ and $p_m^{(m)}$ on $[0,1]$ may be computed in closed form with the results
\begin{equation}\label{eq:pmnorms}
	\|p_m\|^2 = A_m, \qquad \|p_m'\|^2 = B_m,\qquad
	\|p_m^{(m)}\|^2 = C_m,
\end{equation}
where
\begin{equation}\label{eq:ABC}
	A_m= \frac{1}{2} - \frac{(2m)!^4}{4(4m)!m!^4},\quad B_m=\frac{(2m-2)!^2(2m-1)!^2}{(4m-3)!(m-1)!^4},\quad
	C_m =  \frac{(2m-2)!(2m-1)!}{(m-1)!^2}.
\end{equation} 
The first few relevant values are tabulated in Table~\ref{tab:ABC}. Note that $A_m< 1/2$ for all $m$ and also that $0\le p_m(x)\le 1$ for $x\in [0,1]$.
\begin{table}
	\begin{center}\begin{tabular}{|c|c|c|c|c|} \hline
		$m$ & $1$ & $2$ & $3$ & $4$  \\ \hline \hline
		$A_m$ & $1/3$ & $13/35$ & $181/462$ & $521/1287$ \\
		$B_m$ & $1$ & $6/5$ & $10/7$ & $700/429$ \\
		$C_m$ & $1$ & $12$ & $720$ & $100800$ \\ \hline
	\end{tabular}\caption{The first few values of the constants $A_m$, $B_m$ and $C_m$.}
	\label{tab:ABC}
	\end{center}
\end{table}

\subsection{Timelike geodesics}
\label{sub:timelike}

Instead of the timelike convergence condition, we assume that any deviations
from timelike convergence along geodesics may be estimated using Sobolev norms
of the form
\begin{equation}\label{eq:Ricci_hyp}
\int_I  f(t)^2 R_{\mu\nu} U^\mu U^\nu|_{\gamma(t)} \,dt\le |||f|||^2:=Q_m(\gamma) \|f^{(m)}\|^2 + Q_0(\gamma) \|f\|^2,
\end{equation}
for some $m\in\mathbb{N}$ and all smooth real-valued $f$ supported in the interior of the compact interval $I\subset\mathbb{R}$, where $\gamma:I\to M$ is a unit-speed timelike geodesic,
$Q_m(\gamma)$ and $Q_0(\gamma)$ are non-negative constants (independent of $f$) of appropriate dimensions and $\|\cdot\|$ denotes the standard norm of $L^2(I)$. Estimates of this type might arise in solutions to the Einstein equations with matter, due to properties of the matter field, such as QEIs or related constraints, see e.g.,~\cite{Fewster:2006ti,BrownFewsterKontou:2018,Fewster:unpublished}.

Then, assuming that $R_{\mu\nu} U^\mu U^\nu|_{\gamma(t)}$ is continuous on $I$, a simple approximation argument shows that the same bound~\eqref{eq:Ricci_hyp} holds for all $f$ in the Sobolev space $W_0^{m}(I)$, which is the closure of 
$C_0^\infty(\mathrm{int}\, I)$ in the norm $|||\cdot|||$. Each such function $f$
is $m-1$ times continuously differentiable on $I$, with a distributional $m$'th derivative that may be identified with an element of $L^2(I)$, and obeys the generalised Dirichlet conditions $f^{(k)}|_{\partial I}=0$ for $0\le k\le m-1$; see e.g.,~\cite[Thm 4.12(III)]{Adams}. As $I$ is compact, $L^2(I)\subset L^1(I)$ so $f^{(m-1)}$ is the indefinite integral of an element of $L^1(I)$, and is therefore absolutely continuous.
 
Let $S$ be a compact smooth spacelike Cauchy surface in $M$ with extrinsic curvature $K$. Suppose that $\gamma:[0,\tau]\to M$ is a unit speed, future-directed, timelike geodesic emanating normally from $S$ and write $\rho(t)= R_{\mu\nu} U^\mu U^\nu|_{\gamma(t)}$.
By Prop.~\ref{prop:timelike}, $\gamma$ contains a focal point to $S$
if there is a  piecewise smooth $f$ on $[0,\tau]$ with $f(0)=1$ and $f(\tau)=0$, such that
\begin{equation}
J[f]\le -K|_{\gamma(0)}\,,
\end{equation}
where
 \begin{equation}
 J[f]=\int_0^\tau\left( (n-1)\dot{f}(t)^2 + f(t)^2 \rho(t)  \right)\,dt\,.
 \end{equation} 
Our aim is to estimate $J[f]$ for specific functions $f$ defined below,
using~\eqref{eq:Ricci_hyp} to control the contribution of $\rho$. The problem
to be faced is that no function with $f(0)=1$ can belong to $W_0^m([0,\tau])$; we address this by applying~\eqref{eq:Ricci_hyp} to a `rounded off' function and making further assumptions on $\rho$ near $t=0$ to estimate the error incurred. The rounding off could be performed in many ways. We now turn to the two scenarios mentioned above, beginning with the situation in which timelike convergence holds for small positive values of $\tau$.

\subsubsection{Scenario 1} 
\label{subsub:scenario1timelike}
 
Suppose that $\rho(t)= R_{\mu\nu} U^\mu U^\nu|_{\gamma(t)}$ is a smooth function on $[0,\tau]$ that is initially negative, $\rho\le \rho_0\le 0$ on $[0,\tau_0]$ for
some $0<\tau_0<\tau$. That is, timelike convergence (equivalent to the SEC) holds initially, 
with strict inequality if $\rho_0<0$. 

Defining a piecewise smooth function on $[0,\tau]$ by
\begin{equation}\label{eq:scen1fdef}
f(t) = \begin{cases}
1 & t\in [0,\tau_0) \\
I(m,m;(\tau-t)/(\tau-\tau_0)) & t\in[\tau_0,\tau]\,,
\end{cases}
\end{equation}
we note that $f(0)=1$, $f(\tau)=0$. We will prove the following:
\begin{lemma}\label{lem:scenario1}
	For the function $f$ given by~\eqref{eq:scen1fdef} and $\rho$ satisfying~\eqref{eq:Ricci_hyp} on $[0,\tau]$ we have
	\begin{equation}\label{eq:scenario1}
	J[f]\le  \nu_*:= (1-A_m)\rho_0 \tau_0  +
	\frac{Q_m(\gamma) C_m}{\tau_0^{2m-1}} + Q_0(\gamma) A_m\tau  + \frac{(n-1)B_m}{\tau-\tau_0} +
	\frac{Q_m(\gamma) C_m}{(\tau-\tau_0)^{2m-1}}.
	\end{equation}
	Consequently, if $-K|_{\gamma(0)}\ge \nu_*$ then $\gamma$ contains a focal point to $S$.
\end{lemma}
\begin{proof}
The consequences are immediate by Prop.~\ref{prop:timelike}, so it is enough to establish the estimate on $J[f]$. Defining a piecewise smooth function on $[0,\tau]$ by
\begin{equation}\label{eq:scen1phidef}
\varphi(t) = \begin{cases}
I(m,m;t/\tau_0) & t\in [0,\tau_0) \\
1 & t\in[\tau_0,\tau],
\end{cases}
\end{equation}
we note that $f\varphi$ is $m-1$ times continuously differentiable, with  $(\varphi f)^{(m)}$ existing and continuous everywhere except $t=\tau_0$, where it has a finite jump discontinuity. Therefore $\varphi f\in W_0^m([0,\tau])$, and 
writing $f^2=(\varphi f)^2 + (1-\varphi^2)f^2=(\varphi f)^2 + 1-\varphi^2$, we have 
\begin{align}
\int_0^\tau  f(t)^2 \rho(t) \,dt &\le \int_0^{\tau_0}  (1-\varphi(t)^2) \rho(t) \,dt  + 
Q_m(\gamma) \|(\varphi f)^{(m)}\|^2 + Q_0(\gamma) \|\varphi f\|^2\notag \\
&\le \rho_0\int_0^{\tau_0}  (1-\varphi(t)^2)\,dt  + 
Q_m(\gamma) \|(\varphi f)^{(m)}\|^2 + Q_0(\gamma) \|\varphi f\|^2
\end{align}
using~\eqref{eq:Ricci_hyp}, along with the assumption $\rho\le \rho_0$ on $[0,\tau_0]$. Now $\varphi f$ is equal to $\varphi$ on $[0,\tau_0]$ and $f$ on $[\tau_0,\tau]$, and the $L^2$-norms of derivatives of $(\varphi f)$ decompose accordingly. Indeed, using suitably rescaled versions of~\eqref{eq:pmnorms} (see~\eqref{eq:scalednorm}),
\begin{equation}
\|\varphi f\|^2 = A_m \tau_0 + A_m(\tau-\tau_0)= A_m\tau,\qquad
\|(\varphi f)^{(m)}\|^2 = \frac{C_m}{\tau_0^{2m-1}} +  \frac{C_m}{(\tau-\tau_0)^{2m-1}}  .
\end{equation}
Thus, we have
\begin{equation}
\int_0^\tau  f(t)^2 \rho(t) \,dt \le \rho_0\tau_0 (1-A_m) +  
\frac{Q_m(\gamma)C_m}{\tau_0^{2m-1}} +  \frac{Q_m(\gamma)C_m}{(\tau-\tau_0)^{2m-1}} + Q_0(\gamma) A_m\tau
\end{equation}
and since $\|f'\|^2 = B_m/(\tau-\tau_0)$, the estimate on $J[f]$ is complete.
\end{proof}

Now using Propositions \ref{prop:timelike} and \ref{prop:timelikeincom} the following theorem is immediate. 
\begin{theorem}
	\label{the:sing}
	Let $(M,g)$ be a smooth globally hyperbolic spacetime of dimension $n >2$ and let $S$ be a smooth compact spacelike Cauchy surface in $M$. Suppose that, for some $\tau>0$, there is an integer $m\ge 1$ and constants $Q_m$ and $Q_0$ so that: 
	\begin{itemize}
		\item[(i)] the Ricci tensor obeys~\eqref{eq:Ricci_hyp} along every
		unit-speed future-directed timelike geodesic $\gamma$ of length $\tau$ emanating normally from $S$, with $Q_m(\gamma)\le Q_m$, $Q_0(\gamma)\le Q_0$;  
		\item[(ii)] there exists $\rho_0\le 0$ and $\tau_0\in (0,\tau)$ so that along every such geodesic, $R_{\mu \nu} \dot{\gamma}^\mu \dot{\gamma}^\nu |_{\gamma(t)}\le \rho_0$ for $t\in [0,\tau_0]$;
		\item[(iii)] the initial extrinsic curvature of $S$ satisfies
			\be
			-K  \geq \min\left\{\frac{(n-1)}{\tau_0} ,\nu_* \right\}
			\ee			
			everywhere on $S$, where $\nu_*$ is given by~\eqref{eq:scenario1},
			but with $Q_k(\gamma)$ replaced by $Q_k$. 
		\end{itemize}
		Then no future-directed timelike curve emanating from $S$ has length greater than $\tau$ and $M$ is future timelike geodesically incomplete. 
\end{theorem} 
We conclude this section with two remarks. First,
since a focal point in $[0,\tau']$, for $\tau'<\tau$ certainly implies the existence of a focal point in $[0,\tau]$, we can replace $\nu_*$ by 
\[
\nu_{**}: = (1-A_m)\rho_0 \tau_0  +
\frac{Q_m C_m}{\tau_0^{2m-1}} +
\min_{\tau'\in (\tau_0,\tau]}\left(
Q_0 A_m\tau'  + \frac{(n-1)B_m}{\tau'-\tau_0} +
\frac{Q_m C_m}{(\tau'-\tau_0)^{2m-1}}\right)
\]
in the statement of Lemma~\ref{lem:scenario1} and Theorem~\ref{the:sing} if we wish. Similarly,
If $\gamma:[0,\infty)\to M$ is future complete and $-K|_{\gamma(0)}$ is greater than or equal to 
the minimum value of $\nu_*$ over $\tau\in (\tau_0,\infty)$ then $\gamma$ contains a focal point to $S$.

Second, if $(n-1)/\tau_0\le \nu_*$ on $S$ then our result simply reduces to the original Hawking singularity theorem, and no future-directed timelike curve emanating from $S$ has length greater than $\tau_0$, as in 
Proposition \ref{prop:timelikeincom} and Corollary \ref{cor:timelikeincom}).
Therefore it is important to show that there are situations in which  $\nu_*<(n-1)/\tau_0$, so that our result represents a nontrivial improvement. 

For this purpose, it is convenient to write $\tau=(1+\mu B_m)\tau_0$ for $\mu>0$, whereupon we may discard the term containing $\rho_0$ in the definition of $\nu_*$ to give
\begin{equation}
	\nu_* \le \frac{Q_m C_m}{\tau_0^{2m-1}}+  Q_0 A_m\tau_0  + 
	\frac{n-1}{\mu\tau_0}  + Q_0 A_m B_m\tau_0 \mu + \frac{Q_m C_m}{(B_m\tau_0)^{2m-1}\mu^{2m-1}}. 
\end{equation}
Clearly $\nu_*\ge (n-1)/\tau_0$ if $\mu\le 1$, so we 
restrict to the case $\mu>1$ and use $1/\mu^{2m-1}<1/\mu$, obtaining
\begin{equation}\label{eq:nuEFG}
	\nu_*\tau_0 \le  E  + F\mu + \frac{G}{\mu}\,,
\end{equation}
where
\begin{equation}
	E = \frac{Q_m C_m}{\tau_0^{2(m-1)}}+  A_m Q_0\tau_0^2 ,\qquad
	F=A_m B_m Q_0 \tau_0^2 ,\qquad 
	G = n-1 + \frac{Q_m C_m}{B_m^{2m-1}\tau_0^{2(m-1)}}. 
\end{equation}
As the right-hand side of~\eqref{eq:nuEFG} takes its minimum at
$\mu = \sqrt{G/F}$, it follows that
\begin{equation}
\nu_*\tau_0 \le \begin{cases} E+2\sqrt{FG} &  F\le G\\  E+F+G & G<F
\end{cases}
\end{equation}
(the second case is uninteresting because $G>n-1$). 
Accordingly, the conditions 
\begin{equation}
	F\le G\quad\text{and}\quad E + 2\sqrt{FG}<n-1
\end{equation}
imply $\nu_*\tau_0<n-1$ 
(evidently this can only be satisfied if $F<(n-1)/4$). In particular, 
it holds if $Q_0\tau_0^2 \ll 1$ and $Q_m/\tau_0^{2(m-1)} \ll 1$, in which case the
optimising value of $\mu$ is much larger than unity, 
\begin{equation}
	\mu\sim \sqrt{\frac{n-1}{A_mB_m Q_0\tau_0^2}}
\end{equation}
leading to the prediction of a focal point within a timescale 
\begin{equation}
	\label{eqn:timescale}
	\tau\sim \sqrt{\frac{(n-1)B_m}{A_m Q_0}},
\end{equation}
with initial extrinsic curvature obeying
\begin{equation}
	\label{eqn:extcurv}
		-K|_{\gamma(0)} >  \nu_*\sim \frac{n-1}{\tau_0}\sqrt{\frac{4 A_m B_m Q_0\tau_0^2}{n-1} }\ll \frac{n-1}{\tau_0}
\end{equation}
(assuming that $E$ and $F$ are of comparable order, so that $E\ll \sqrt{F}$).
Importantly, this level of extrinsic curvature is not sufficient to guarantee a focal point within time $\tau_0$, the period in which we have assumed timelike convergence to hold. In this regime, we also have
\begin{equation}
\nu_*\sim \frac{2(n-1)B_m}{\tau},
\end{equation}
which is of the same order of magnitude as the initial contraction that would be needed to guarantee a focal point in $(0,\tau]$ assuming timelike convergence (see Cor.~\ref{cor:timelikefocal}). Therefore our result provides a Hawking-type singularity theorem with weakened energy conditions, but without drastically increased contraction.

\subsubsection{Scenario 2}  
Let us now drop the assumption that timelike convergence holds for small positive times along geodesics
leaving $S$. Starting with a single unit-speed timelike geodesic $\gamma:[0,\tau]\to M$, with $\gamma(0)\in S$, extend $\gamma$ backwards in time, still as a unit-speed timelike geodesic, to obtain $\gamma:[-\tau_0,\tau]\to M$. We assume that~\eqref{eq:Ricci_hyp} holds on the extended geodesic, for all $f\in W_0^m([-\tau_0,\tau])$ obeying generalised Dirichlet boundary conditions at $t=-\tau_0,\tau$. 

As in Scenario 1, we aim to estimate $J[f]$ for a piecewise smooth function $f$ on $[0,\tau]$ obeying $f(0)=1$, $f(\tau)=0$, namely
\begin{equation}\label{eq:scen2fdef}
f(t) = \begin{cases} I(m,m;(\tau'-t)/\tau') & t\in [0,\tau'] \\ 0 & t\in [\tau',\tau],
\end{cases}
\end{equation}
where the free parameter $\tau'\in [0,\tau]$. Defining
\begin{equation}\label{eq:Hhat1}
\hat{L}_1(\tau')=\frac{Q_m(\gamma) C_m}{(\tau')^{2m-1}}  + \frac{(n-1)B_m}{\tau'} + A_m Q_0(\gamma) \tau'
\end{equation}
and
\begin{equation} \label{eq:Hhat2}
\hat{L}_2(\tau_0') = 
\frac{Q_m(\gamma) C_m}{(\tau_0')^{2m-1}}  + A_m (Q_0(\gamma)-\rho_{\min}) \tau_0' \,,
\end{equation}
and also
\begin{equation}\label{eq:H12}
L_1(\tau) = \min_{\tau'\in (0,\tau]} \hat{L}_1(\tau') ,\qquad
L_2(\tau_0) = \min_{\tau_0'\in (0,\tau_0]}\hat{L}_2(\tau_0')\,,
\end{equation}
 we prove: 
\begin{lemma}\label{lem:scenario2}
	For the function $f$ given by~\eqref{eq:scen2fdef}, and $\rho$ satisfying~\eqref{eq:Ricci_hyp} on $[-\tau_0,\tau]$ we have
\begin{equation}
J[f]\le \hat{L}_1(\tau') + L_2(\tau_0) \,.
\end{equation}
Consequently, if $-K|_{\gamma(0)}\ge  L_1(\tau) + L_2(\tau_0)$
then there is a focal point to $S$ along $\gamma$ in the proper time interval $[0,\tau]$.  
\end{lemma}
\begin{proof}
We have
\begin{equation}
J[f] = \frac{(n-1)B_m}{\tau'} + \int_0^{\tau'}  f^2 R_{\mu\nu} U^\mu U^\nu\,dt \,.
\end{equation}
To estimate the integral, we first extend the domain of $f$ to create a function $\tilde{f}\in W_0^m([-\tau_0,\tau])$ that agrees with $f$ on $[0,\tau]$ and is given by 
\begin{equation}
\tilde{f}(t) = \begin{cases}  0 & t\in [-\tau_0,-\tau_0'] \\ I(m,m;-t/\tau_0') & t\in [-\tau_0',0),
\end{cases}
\end{equation}
on $[-\tau_0,0)$, where $\tau_0'\in(0,\tau_0]$ is arbitrary. We then estimate
\begin{align}
\int_0^{\tau'}  f^2 R_{\mu\nu} U^\mu U^\nu\,dt &= 
\int_{-\tau_0}^{\tau}  \tilde{f}^2 R_{\mu\nu} U^\mu U^\nu\,dt - \int_{-\tau_0'}^{0}  \tilde{f}^2 R_{\mu\nu} U^\mu U^\nu\,dt \\
&\le Q_m(\gamma) \|\tilde{f}^{(m)}\|^2 + Q_0(\gamma)\|\tilde{f}\|^2 -\rho_{\min} \int_{-\tau_0'}^{0}  \tilde{f}^2 \,dt
\end{align}
where the norms are those of $L^2(-\tau_0,\tau)$ and the constants $Q_k(\gamma)$ refer to the extended geodesic, while $\rho_{\min} = \min_{[-\tau_0,0]} \rho$. The right-hand side may be evaluated by using our results on incomplete Beta functions, giving an overall upper bound
\begin{equation}
 J[f] \le \hat{L}_1(\tau') + \hat{L}_2(\tau_0')
\end{equation}
and the first part of result follows on using the freedom to optimise over $\tau_0'$. The second is proved using Prop.~\ref{prop:timelike}, and optimising over $\tau'\in (0,\tau]$.
\end{proof}
 
The minimum used to define $L_2$ may be found explicitly, giving
\begin{equation}
L_2(\tau_0) =  
\frac{2mQ_m(\gamma)}{2m-1} ((Q_0(\gamma)-\rho_{\min})A_m)^{1-1/(2m)} ((2m-1)C_m)^{1/(2m)}
\end{equation}
for $Q_0(\gamma)-\rho_{\min} > (2m-1)Q_m(\gamma)C_m/(A_m\tau_0^{2m})$, and 
\begin{equation}
L_2(\tau_0) =  
\frac{Q_m(\gamma) C_m}{(\tau_0)^{2m-1}}  + A_m (Q_0(\gamma)-\rho_{\min}) \tau_0 
\end{equation}
otherwise. A closed form expression for $L_1(\tau)$ is not possible for general $m$. However, one may note that $\hat{L}_1$ has a single minimum in $(0,\infty)$ and no other critical points. As the only positive contribution to $\hat{L}_1'$ arises from the last term, it follows that if $Q_0\tau^2\le (n-1)B_m/A_m$ then  the gradient at $\tau$ is negative and $L_1(\tau)=\hat{L}_1(\tau)$.

By analogy with Theorem~\ref{the:sing} we now have:
\begin{theorem}
	\label{the:singscen2}
	Let $(M,g)$ be a smooth globally hyperbolic spacetime of dimension $n >2$ and let $S$ be a smooth compact spacelike Cauchy surface in $M$. Suppose that, for some $\tau,\tau_0>0$ there is an integer $m\ge 1$ and constants $Q_m$ and $Q_0$ so that
	\begin{itemize}
		\item[(i)] every unit-speed future-directed timelike geodesic of length $\tau$ emanating normally from $S$ can be extended as an affine geodesic to
		$\gamma:[-\tau_0,\tau]\to M$ with $\gamma(0)\in S$, along which the Ricci tensor obeys~\eqref{eq:Ricci_hyp} with $Q_m(\gamma)\le Q_m$, $Q_0(\gamma)\le Q_0$; 
		\item[(ii)] there exists a finite lower bound $\rho_{\min}$
		so that $R_{\mu \nu} \dot{\gamma}^\mu \dot{\gamma}^\nu |_{\gamma(t)}\ge \rho_{\min}$ for every such geodesic and all $t\in [-\tau_0,0]$;
		\item[(iii)] the extrinsic curvature of $S$ satisfies
		\be
		-K  \geq  L_1(\tau) + L_2(\tau_0)
		\ee			
		everywhere on $S$, where the functions $L_i$ are given by~\eqref{eq:Hhat1},~\eqref{eq:Hhat2} and~\eqref{eq:H12}, with $Q_k(\gamma)$ replaced by $Q_k$ ($k=0,m$).   
	\end{itemize}
	Then no future-directed timelike curve emanating from $S$ has length greater than $\tau$ and $M$ is future timelike geodesically incomplete. 
\end{theorem} 

Under the assumption of timelike convergence, the conclusions of this result hold provided the initial contraction satisfies $-K\ge (n-1)/\tau$ on $S$ (Cor.~\ref{cor:timelikefocal} and Prop.~\ref{prop:timelikeincom}). It is important to note that there are circumstances in which the contraction required by Theorem~\ref{the:singscen2} is of a similar order. 

For example, suppose that $\rho_{\min}\ge 0$ and the following conditions hold 
\begin{equation}
Q_0\tau^2\ll B_m/A_m, \qquad \tau_0\ll \tau\lesssim \frac{\tau_0^{2m-1}}{Q_m C_m}\,.
\end{equation}
Then certainly $Q_0\tau^2\le B_m/A_m$, so $L_1(\tau)=\hat{L}_1(\tau)$ as noted above. As $Q_mC_m/\tau^{2(m-1)}\lesssim (\tau_0/\tau)^{2(m-1)}\ll 1$ (for $m>1$) we find that $L_1(\tau)=\hat{L}(\tau)\sim (n-1)B_m/\tau$. In the case $m=1$ the assumptions imply $Q_1\ll 1$ and hence $Q_1C_1/\tau\ll (n-1)/\tau$, giving 
the same conclusion for $L_1(\tau)$ as before. 
Turning to $L_2$, the requirement that $\rho_{\min}\ge 0$ implies
\begin{equation}
L_2(\tau_0)\le \hat{L}_2(\tau_0) \le  
\frac{Q_m C_m}{(\tau_0)^{2m-1}}  + A_m Q_0\tau_0
\end{equation}
Under our assumptions, the second term is small relative to $B_m/\tau$, while the first is $\lesssim 1/\tau$. Overall, 
\begin{equation}
	\label{eq:scen2approx}
L_1(\tau)+L_2(\tau_0)\lesssim \frac{(n-1)B_m+1}{\tau}\,,
\end{equation}  
which is of comparable order to $(n-1)/\tau$ at least for small values of $m$. Thus the 
contraction required by Theorem~\ref{the:singscen2} is comparable to $(n-1)/\tau$. 

A feature of Theorem~\ref{the:singscen2} is that increasing $\rho_{\min}$ --- the extent to which timelike convergence fails at small negative times --- decreases the required initial contraction. This is because the QEI type constraints require that a period in which timelike convergence fails is followed shortly by a period in which timelike convergence is satisfied even more strongly. This is analogous to the quantum interest effect~\cite{FordRoman:1999,FewsterTeo:2000,Fewster:2010gm,BrownFewsterKontou:2018}. If $\rho_{\min} \leq 0$ then Theorem~\ref{the:singscen2} typically overestimates the required contraction and scenario 1 should be used instead.

\subsection{Null geodesics}

Let $\gamma$ be any null geodesic in $M$. Using the same invariant notation as in Prop.~\ref{prop:null}, we suppose that the Ricci tensor obeys   
\begin{equation}\label{eq:Ricci_hypnull}
\int_\gamma  f^2 \mathrm{Ric}(d\gamma,d\gamma)  \le |||f|||^2:=Q_m(\gamma;w) \|w^{1-m} \nabla_{d\gamma}^m f\|^2 +  Q_0(\gamma;w) \|w f\|^2,
\end{equation}
for every smooth compactly supported $(-\tfrac{1}{2})$-density $f$ on $\gamma$,
and every choice of smooth positive density $w$ on $\gamma$, so that
$Q_k(\gamma;\lambda w)=\lambda^{2(k-1)}Q_k(\gamma;w)$ for any $\lambda>0$.
Here $\|\cdot\|$ is the intrinsic $L^2$-norm on $\tfrac{1}{2}$-densities defined on $\gamma$, and appropriate powers of the density $w$ are inserted to ensure that the arguments of these norms have the correct weight as densities.

Now let $P$ be a future converging spacelike submanifold of $M$ of co-dimension $2$ with mean normal curvature vector field $H^\mu$. Suppose that $\gamma$ is a 
future-directed null geodesic emanating normally from $P$. Extending $\hat{H}_\mu$ by parallel transport along $\gamma$, we define a density $w=\hat{H}_\mu d\gamma^\mu_+/d\lambda$. Next choose an affine coordinate $\lambda$ on $\gamma$, so that $p=\gamma(0)$ and $\hat{H}_\mu d\gamma^\mu/d\lambda =1$. Then $w=1$, $q=\gamma(\ell)$, with $\ell=L_{\hat{H}}(\gamma)$ and Eq.~\eqref{eq:Ricci_hypnull} becomes
\be
\label{eqn:Riccinull}
\int_0^\ell f(\lambda)^2 R_{\mu \nu} U^\mu U^\mu d\lambda \leq Q_m(\gamma) \|  f^{(m)} \|^2+Q_0 (\gamma) \| f\|^2 \,.
\ee
Writing $\rho(\lambda)= R_{\mu\nu} U^\mu U^\nu|_{\gamma(\lambda)}$, and by Prop.~\ref{prop:null}, there is a focal point to $P$ along $\gamma$ if there is a  piecewise smooth $f$ on $[0,\ell]$ with $f(0)=1$ and $f(\ell)=0$, such that
\begin{equation}
	J[f]\le -(n-2)H |_{\gamma(0)}\,,
\end{equation}
where
\begin{equation}
	J[f]=\int_0^\ell \left( (n-2)f'(\lambda)^2 + f(\lambda)^2 \rho(\lambda)  \right)\,d\lambda\,.
\end{equation} 

The two scenarios following are analogous to the ones in Sec.~\ref{sub:timelike}. In the first, we will suppose that the NEC is satisfied at small positive values of $\lambda$ while in the second, we impose conditions on $\rho$ at small negative values of $\lambda$. 
 
\subsubsection{Scenario 1}

Suppose that initially the NEC is satisfied, and so let $\rho(\lambda)= R_{\mu\nu} U^\mu U^\nu|_{\gamma(\lambda)}$ be a smooth function on $[0,\ell]$ that is initially negative, $\rho \le \rho_0\le 0$ on $[0,\ell_0]$ for some $0<\ell_0<\ell$. 

Defining $f:[0,\ell]\to\mathbb{R}$ by~\eqref{eq:scen1fdef} with $\ell$ instead of $\tau$ and $\ell_0$ instead of $\tau_0$, we can estimate $J[f]$ following the exact steps of the proof of Lemma~\ref{lem:scenario1}. The following Lemma is immediate using Prop.~\ref{prop:null}.
\begin{lemma}\label{lem:scenario1null}
	For $\rho$ satisfying~\eqref{eqn:Riccinull} on $[0,\ell]$ we have
	\begin{equation}\label{eq:scenario1null}
		J[f]\le  \nu_*:= (1-A_m)\rho_0 \ell_0  +
		\frac{Q_m C_m}{\ell_0^{2m-1}} + Q_0 A_m\ell  + \frac{(n-2)B_m}{\ell-\ell_0} +
		\frac{Q_m C_m}{(\ell-\ell_0)^{2m-1}}.
	\end{equation}
	Consequently, if $-(n-2)H|_{\gamma(0)}\ge \nu_*$ then $\gamma$ contains a focal point to $P$. 
\end{lemma}
There is an obvious adaptation of this result to future-complete null geodesics by minimising over $\ell\in (\ell_0,\infty)$. Using Propositions \ref{prop:null} and \ref{prop:nullincompl} the following theorem is immediate. 
\begin{theorem}
	\label{the:singnull}
	Let $M$ be globally hyperbolic with non-compact Cauchy hypersurfaces and let $P$ be a compact achronal future converging spacelike submanifold of $M$ of co-dimension $2$ with mean normal curvature vector $H^\mu=H \hat{H}^\mu$ where $\hat{H}^\mu$ is a future-pointing timelike unit vector. Suppose that for some $\ell>0$ there is an integer $m\ge 1$ and constants $Q_m$ and $Q_0$ so that:  
	\begin{itemize}
		\item[(i)] the Ricci tensor obeys~\eqref{eq:Ricci_hypnull} along every
		future-directed null geodesic $\gamma$ of $\hat{H}$-length $\ell$ emanating normally from $P$, with $Q_m(\gamma)\le Q_m$, $Q_0(\gamma)\le Q_0$; 
		\item[(ii)] there exists $\rho_0\le 0$ and $\ell_0\in (0,\ell)$ so that along every such geodesic $\gamma$, the inequality $\mathrm{Ric}(\gamma',\gamma')\le \rho_0 g(d\gamma,\hat{H})^2$ holds along the initial portion of $\gamma$ with $\hat{H}$-length $\ell_0$;
		\item[(iii)] the mean normal curvature of $P$ satisfies
		\be
		-(n-2)H  \geq \min\left\{\frac{n-2}{\ell_0} ,\nu_* \right\}
		\ee			
		everywhere on $P$, where $\nu_*$ is given by~\eqref{eq:scenario1null}. 
	\end{itemize}
	Then there is an inextendible future-directed null geodesic emanating from $P$ with $\hat{H}$-length less than $\ell$. In particular, $M$ is future null geodesically incomplete. 
\end{theorem} 

Similarly to the timelike version if $(n-2)/\ell_0\le \nu_*$ on $P$ our result reduces to the original Penrose singularity theorem as in Proposition \ref{prop:nullincompl}. So we can examine situations in which $(n-2)/\ell_0\ge \nu_*$ following the exact similar analysis as in \ref{subsub:scenario1timelike} but again with with $\lambda$ instead of $t$, $\ell$ instead of $\tau$, $\ell_0$ instead of $\tau_0$ and Eq.~\eqref{eqn:Riccinull} instead of \eqref{eq:Ricci_hyp}. 

Then for $Q_0\ell_0^2 \ll 1$ and $Q_m/\ell_0^{2(m-1)} \ll 1$ there is a focal point before parameter
\begin{equation}
	\label{eqn:ell}
	\ell \sim \sqrt{\frac{(n-2)B_m}{A_m Q_0}},
\end{equation}
if the magnitude of the mean curvature vector of $P$ satisfies
\begin{equation}
	\label{eqn:meancurv}
	-(n-2)H >  \nu_*\sim \sqrt{4 A_m B_m Q_0(n-2)} \sim \frac{2B_m(n-2)}{\ell}\,.
\end{equation}
Again, this is of the same order as the mean curvature needed to guarantee a focal point if the null convergence condition held (Cor.~\ref{cor:penrfocal}).

\subsubsection{Scenario 2}

In Scenario 2 we drop the assumption that the NEC holds. We instead extend the previously defined $\gamma$ to negative values of $\lambda$ to get $\gamma:[-\ell_0,\ell]\to M$. Next we assume that Eq.~\eqref{eqn:Riccinull} holds on the extended geodesic, for all $f\in W_0^m([-\ell_0,\ell])$ obeying generalised Dirichlet boundary conditions at $\lambda=-\ell_0,\ell$. 

Similarly to the timelike case we want to estimate $J[f]$ and we can follow the same analysis with $\lambda$, $\ell$ and $\ell_0$ instead of $t$, $\tau$ and $\tau_0$. Set

\begin{equation}\label{eq:Hhat1null}
\hat{L}_1(\ell')=\frac{Q_m(\gamma) C_m}{(\ell')^{2m-1}}  + \frac{(n-2)B_m}{\ell'} + A_m Q_0(\gamma) \ell'
\end{equation}
and
\begin{equation} \label{eq:Hhat2null}
\hat{L}_2(\ell_0') = 
\frac{Q_m(\gamma) C_m}{(\ell_0')^{2m-1}}  + A_m (Q_0(\gamma)-\rho_{\min}) \ell_0' \,,
\end{equation}
and also 
\begin{equation}\label{eq:H12null}
L_1(\ell) = \min_{\ell'\in (0,\ell]} \hat{L}_1(\ell') ,\qquad
L_2(\ell_0) = \min_{\ell_0'\in (0,\ell_0]}\hat{L}_2(\ell_0')\,.
\end{equation}
Then the proof of the following lemma is analogous to the one for Lemma~\ref{lem:scenario2} where Eq.~\eqref{eqn:Riccinull} is used instead of \eqref{eq:Ricci_hyp}, and finally using Prop.~\ref{prop:null}, optimising over $\ell'\in (0,\ell]$.
\begin{lemma}\label{lem:scenario2null}
	For $\rho$ satisfying~\eqref{eqn:Riccinull} on $[-\ell_0,\ell]$ we have
	\begin{equation}
		J[f]\le \hat{L}_1(\ell') + L_2(\ell_0).
	\end{equation} 
	Consequently, if $-(n-2) H\ge  L_1(\ell) + L_2(\ell_0)$
	then there is a focal point to $P$ along $\gamma$ in $[0,\ell]$.  
\end{lemma} 
More explicitly, we have
\begin{equation}
	L_2(\ell_0) =  
	\frac{2mQ_m(\gamma)}{2m-1} ((Q_0(\gamma)-\rho_{\min})A_m)^{1-1/(2m)} ((2m-1)C_m)^{1/(2m)}
\end{equation}
for $Q_0(\gamma)-\rho_{\min} > (2m-1)Q_m(\gamma)C_m/(A_m\ell_0^{2m})$, and 
\begin{equation}
	L_2(\ell_0) =  
	\frac{Q_m(\gamma) C_m}{(\ell_0)^{2m-1}}  + A_m (Q_0(\gamma)-\rho_{\min}) \ell_0 
\end{equation}
otherwise. Our main result in this scenario is: 
\begin{theorem}
	\label{the:singscen2null}
		Let $M$ be globally hyperbolic with non-compact Cauchy hypersurfaces and let $P$ be a compact achronal future converging spacelike submanifold of $M$ of co-dimension $2$ with mean normal curvature vector $H^\mu$. 
		Suppose that for some $\ell,\ell_0>0$ there is an integer $m\ge 1$ and constants $Q_m$ and $Q_0$ so that:   
	\begin{itemize}
		\item[(i)] every future-directed null geodesic of $\hat{H}$-length $\ell$ emanating normally from $P$ may be extended to the past, to give a geodesic $\gamma$ with $L_{\hat{H}}(\gamma)=\ell+\ell_0$ so that 
		the Ricci tensor obeys~\eqref{eq:Ricci_hypnull} along $\gamma$ 
		with $Q_m(\gamma)\le Q_m$, $Q_0(\gamma)\le Q_0$; 
		\item[(ii)] there exists a finite lower bound $\rho_{\min}$
		so that $\mathrm{Ric}(d\gamma,d\gamma) \ge \rho_{\min} g(\hat{H},d\gamma)^2$ on the portion of each such geodesic $\gamma$ to the past of $P$; 
		\item[(iii)] the mean normal curvature of $P$ satisfies
		\be
		-(n-2)H  \geq  L_1(\ell) + L_2(\ell_0)\,,
		\ee			
		 where the functions $L_i$ are given by~\eqref{eq:Hhat1null},~\eqref{eq:Hhat2null} and~\eqref{eq:H12null}, with $Q_k(\gamma)$ replaced by $Q_k$ ($k=0,m$).   
	\end{itemize}
	Then there is an inextendible future-directed null geodesic emanating from $P$ with $\hat{H}$-length less than $\ell$. In particular, $M$ is future null geodesically incomplete.  
\end{theorem} 

Similarly to the timelike case it is important to note that there are circumstances in which the mean normal curvature required by Theorem~\ref{the:singscen2null} is of the same order as that in  Corollary~\ref{cor:penrfocal}. 

Suppose that $\rho_{\min}\ge 0$ and the following conditions hold 
\begin{equation}
	Q_0\ell^2\ll B_m/A_m, \qquad \ell_0\ll \ell\lesssim \frac{\ell_0^{2m-1}}{Q_m C_m}\,.
\end{equation}
Then $Q_0\ell^2\le B_m/A_m$, so $L_1(\ell)=\hat{L}_1(\ell)$. Similar to the timelike case  $L_1(\ell)=\hat{L}(\ell)\sim (n-2)B_m/\ell$. 

Turning to $L_2$ we have
\begin{equation}
	L_2(\ell_0)\le \hat{L}_2(\ell_0) \le  
	\frac{Q_m C_m}{(\ell_0)^{2m-1}}  + A_m Q_0\ell_0 \,,
\end{equation}
and overall 
\begin{equation}
	\label{eqn:scen2estnull}
	L_1(\ell)+L_2(\ell_0)\lesssim \frac{(n-2)B_m+1}{\ell}\,,
\end{equation}  
which is of comparable order to $(n-2)/\ell$ at least for small values of $m$. 

If $\rho_{\min} \leq 0$ then Theorem~\ref{the:singscen2null} typically overestimates the required $H$ and scenario 1 should be used.

\section{Applications to the Einstein-Klein-Gordon theory}
\label{sec:applications}

In this section we apply the results of Sec.~\ref{sec:QEI} to the non-minimally coupled classical Einstein-Klein-Gordon theory. The classical non-minimally coupled scalar field is a famous example that violates both the SEC and the NEC, thus lying outside the scope of the original singularity theorems. Singularity theorems with weakened energy conditions  were proven for the null case in \cite{Fewster:2010gm} and for the timelike case in \cite{BrownFewsterKontou:2018}. Here we calculate the required initial contraction for singularity formation using the method described in previous sections and compare our results with Ref.~\cite{Fewster:2010gm} and \cite{BrownFewsterKontou:2018}. This will help illustrate the application of our method and its advantages compared to methods using the Raychaudhuri equation. 

Let $(M,g, \phi)$ be a solution to the Einstein-Klein-Gordon equation in $n>2$ spacetime dimensions, i.e., $G_{\mu\nu}=-8\pi T_{\mu\nu}$, where the right-hand side is the stress-energy tensor
\be
\label{eqn:tmunu}
T_{\mu \nu}=(\nabla_\mu \phi)(\nabla_\nu \phi)+\frac{1}{2} g_{\mu \nu} (m^2 \phi^2-(\nabla \phi)^2)+\xi(g_{\mu \nu} \Box_g-\nabla_\mu \nabla_\nu-G_{\mu \nu}) \phi^2 \,,
\ee
for the nonminimally coupled scalar field $\phi$ obeying the equation 
\be
(\Box +\lambda^{-2} +\xi R)\phi=0 \,,
\ee
where $\lambda$ is a fixed characteristic wavelength and $\xi$ the coupling constant. We suppose that $\xi \in [0,\xi_c]$ where $\xi_c= \tfrac{1}{4}(n-2)/(n-1)$ is the value for conformal coupling.

\subsection{The timelike case}
\label{sec:EKGtime}

Consider first the timelike case. Assume that the scalar field magnitude admits global bounds on a timelike geodesic $\gamma$ parametrized by proper time $\tau$
\be
|\phi| \leq \fmax \leq (8\pi \xi)^{1/2}\,, \qquad |\nabla_{\dot{\gamma}} \phi| \leq \fmax' \,.
\ee
Then, on $\gamma$, and following Theorem 3 and Corollary 1 of Ref.~\cite{BrownFewsterKontou:2018}, we have that
\be
\label{eqn:ekgencon}
\int_\gamma R_{\mu \nu} \dot{\gamma}^\mu \dot{\gamma}^\nu f(\tau)^2d\tau \leq Q(\| \dot{f} \|^2+\tilde{Q}^2 \| f\|^2) \,,
\ee
for any real valued function $f$ of compact support, with constants $Q$ and $\tilde{Q}$ given by
	\be\label{eq:QQtdef}
Q=\frac{32\pi \xi \fmax^2}{1-8\pi \xi \fmax^2} \,, \qquad \tilde{Q}^2=\frac{(1-2\xi)\lambda^{-2}}{4\xi(n-2)}+\left(\frac{8\pi \xi \fmax \fmax'}{1-8\pi \xi \fmax^2}\right)^2  \,.
\ee 
Eq.~\eqref{eqn:ekgencon} is of the form of Eq.~\eqref{eq:Ricci_hyp} with $m=1$, $Q_1=Q$ and $Q_0=Q\tilde{Q}^2$. To estimate the values of these constants we first reinsert the units and restore the constants $G$ and $c$
\be\label{eq:QQtdefdim}
Q=\frac{32\pi \xi G\fmax^2/c^4}{1-8\pi \xi G\fmax^2/c^4} \,, \qquad \tilde{Q}^2=\frac{(1-2\xi)\lambda^{-2}c^2}{4\xi(n-2)}+\left(\frac{8\pi \xi G\fmax \fmax'/c^4}{1-8\pi \xi G\fmax^2/c^4}\right)^2  \,.
\ee 

It is interesting to estimate these values in a situation where $\lambda$ is the reduced Compton wavelength of an elementary particle, making appropriate choices for $\fmax$ and $\fmax'$. 
In Ref.~\cite{BrownFewsterKontou:2018} the value of $\fmax$ was estimated by considering a quantized scalar field in Minkowski spacetime of dimension $n$, in a thermal state of temperature $T<T_\lambda$, where $T_\lambda=c \hbar/(\lambda k)$ is the reduced Compton temperature of the particle and $k$ is Boltzmann's constant. The temperature $T_\lambda$ defines a scale beyond which the model cannot be trusted. With these considerations $\fmax^2 \sim \langle {:}\phi^2{:}\rangle_T$ and 
\begin{equation}
	Q\sim  (\ell_{\textrm{Pl}}/\lambda)^{n-2}(T/T_\lambda)^{(n-2)/2} K_{(n-2)/2}(T_\lambda/T) \,,
\end{equation}
as was shown in Ref.~\cite{BrownFewsterKontou:2018} (in that reference the mass was used instead of the reduced Compton length). Here $K_\nu$ is a modified Bessel function of the second kind and $\ell_{\textrm{Pl}}$ is the Planck length. 

From Eq.~\eqref{eq:QQtdefdim} the second term of $\tilde{Q}^2$ is proportional to $Q^2 (\fmax'/\fmax)^2$. On dimensional grounds the ratio $\fmax'/\fmax$ is proportional to $c/\lambda$. For $\lambda$ taken as the Compton length of elementary particles $Q\ll 1$, as will be seen shortly. Therefore the first term of $\tilde{Q}^2$ is expected to be much larger than the second and so $\tilde{Q} \sim c/\lambda$. 

In order to give a quantitative, and partly heuristic, illustration of our results, we consider the following toy model. Suppose the universe were described by a Einstein--Klein--Gordon model, in which the characteristic length scale of the scalar is the reduced Compton wavelength of a pion. Given an expansion rate of the universe, and other conditions, drawn from actual cosmological data, would one be able to conclude that the universe is necessarily past timelike geodesically incomplete? To do this we must consider the time-reverse of the analysis presented above, so the question is whether the extrinsic curvature of surfaces of constant cosmic time -- the Hubble parameter -- is sufficiently positive; that is, whether the expansion rate is sufficiently large. We bear in mind that the SEC is violated in our actual universe, due to the dominant effect of dark energy, with the ratio of pressure to energy density being very close to $-1$. 
This motivates two different calculations: (a) using parameters drawn from an era in which the SEC did hold, as an instance of Scenario~1, and (b) using parameters corresponding to the present time, and assuming that the SEC will continue to fail for some time $\tau_0$ into the future, as a (time-reversed version of) Scenario~2. 
We will show in each case that, using the pion as the matter model, the expansion rate of the actual universe would be sufficient to conclude past geodesic incompleteness. In fact, there is a caveat to these results, because the values of parameters $Q_0$ and $Q_1$ were derived on the basis that the temperature scale $T$ is not exceeded. This means that what the heuristic argument actually shows is that, on timescales within the age of our actual universe, the toy model universe must display either past geodesic timelike incompleteness on a timescale of the age of our actual universe, or locations where the temperature scales exceed $T$. For brevity, we will describe either of these occurrences as a singularity.

First, we consider Scenario 1, in which the SEC is satisfied for time $\tau_0$. For a neutral pion in $n=4$ dimensions with mass $135\textrm{MeV}/c^2$, we have $\ell_{\textrm{Pl}}/\lambda=1.11 \times 10^{-20}$ and $T_\lambda=1.56 \times 10^{12}\textrm{K}$. (All calculations are made to higher precision but reported to 3S.F.; however it is really the orders of magnitude that are of interest.) Taking $T=10^{-2}T_\lambda$ (corresponding to the temperature of our universe about $1\textrm{s}$ after the big bang) gives an estimate of $Q \sim 5.66 \times 10^{-87}$. Then $Q_1=Q \ll 1$ and $Q_0\sim Q (c/\lambda)^2=2.39\times 10^{-39}\textrm{s}^{-2}$. Thus if $\tau_0$ is of the order of the reduced Compton time ($4.87\times 10^{-24}$s), we have $Q_0 \tau_0^2 \sim Q \ll 1$. In that case the initial contraction is given by Eq.~\eqref{eqn:extcurv} with the units restored
\be\label{eq:nustar}
 \nu_*\sim \lambda^{-1} c \sqrt{12 A_1 B_1 Q} \sim 3.09 \times 10^{-20}\textrm{s}^{-1} \,.
\ee
The maximum allowed temperature for this case is $T=1.56\times 10^{10}\textrm{K}$, while the timescale on which the singularity occurs is given by Eq.~\eqref{eqn:timescale} as
\be\label{eq:piontimescale}
\tau \sim \lambda c^{-1} \sqrt{\frac{3B_1}{A_1}} Q^{-1/2} \sim 1.94 \times 10^{20}\textrm{s} \,,
\ee
or about $6.15 \times 10^{12}$ years. 

For comparison, in our actual universe, and assuming the $\Lambda$CDM model, the SEC was most recently obeyed at time $t_1$ when 
\be
\Omega_\Lambda (t_1)=\frac{\Omega_b (t_1)}{2}+\Omega_r (t_1) \,.
\ee
Here $\Omega_x=\rho_x/\rho_{\text{crit}}$, $\rho_{\text{crit}}$ is the critical energy density, $\Lambda$ corresponds to dark energy, $m$ to matter (baryonic and cold dark matter) and $r$ to radiation. Using the different evolution of each energy density component we find that the redshift when the SEC was last satisfied $z_1$ is given by the solution of 
\be
\Omega_{\Lambda_0}-\frac{\Omega_{m_0}}{2}(z_1+1)^3-\Omega_{r_0} (z_1+1)^4=0 \,,
\ee
where $\Omega_{x0}$ are the respective quantities today. From the most recent results published by the PLANCK collaboration \cite{aghanim2018planck} we have that $\Omega_{\Lambda_0}=0.6889 \pm 0.0056$ and $\Omega_{m_0}=0.311 \pm 0.002$ which gives a redshift of $z_1=0.642$ for when the SEC was last obeyed. From the first Friedmann equation we have 
\be
\frac{H^2(t_1)}{H_0^2}=\Omega_{\Lambda_0}+\Omega_{r_0} (z_1+1)
^4+\Omega_{m_0}(z_1+1)^3 \,,
\ee
where $H(t)$ is the Hubble parameter and $H_0$ its value today. Again from the PLANCK collaboration \cite{aghanim2018planck} we get that $H_0=(2.184 \pm 0.016) \times  10^{-18} \textrm{ s}^{-1}$. That gives $H(t_1)=3.14 \times  10^{-18} \textrm{ s}^{-1}$ at the time when the SEC was last obeyed. This exceeds the minimum threshold \eqref{eq:nustar} by two orders of magnitude, so the toy model universe would necessarily have a past singularity, using these parameters. Partly because the threshold is exceeded by such a margin, the estimated timescale~\eqref{eq:piontimescale} for the location of the singularity is accordingly rather pessimistic. Nonetheless, it is remarkable that it is within two orders of magnitude of the age of the universe at the relevant time.   

If the characteristic scale is replaced by the reduced Compton wavelength of 
more massive particles, the same calculations produce higher expansion thresholds beyond which a singularity is inevitable. For a proton, with mass $938\textrm{MeV}/c^2$, and a maximum  temperature $T=1.09\times 10^{11}\textrm{K}$, the threshold is $1.50\times 10^{-18}s^{-1}$, which is still marginally exceeded by the measured value. The timescale for the singularity is now $1.27\times 10^{11}$ years.
On the other hand, for a Higgs particle, with mass $125\textrm{GeV}/c^2$ and with $T=1.44 \times 10^{13}\textrm{K}$ (the temperature of the Universe at age $10^{-4}\textrm{s}^{-1}$), the threshold is
$\nu_*=2.68 \times 10^{-14}\textrm{s}^{-1}$. This threshold is larger than the observed Hubble parameter by $4$ orders of magnitude, so our results would be inconclusive in that case. 
These illustrations are intended purely to show that the results we have obtained are quantitative and capable of producing plausible cosmological results in the toy models for the thresholds beyond which a singularity is inevitable, and the timescales on which they occur. 

Moving to Scenario 2 where the requirement that the SEC holds is dropped, we are taking into account the behaviour of $\rho$ just \emph{after} the time $\tau=0$ at which the Hubble parameter is measured (recall that we are using time-reversed versions of our earlier results). Here we assume that the SEC is violated for the time interval $[0,\tau_0]$, and $\rho_{\textrm{min}} >0$ which as we mentioned is compatible with current cosmological observations. 

First we observe that for the pion  $Q_0^{-1/2}\sim 6.49 \times 10^{11}$ years is large in comparison to the lifetime of the universe. Then $Q_0 \tau^2 \ll B_1/A_1$ for focal points occurring within a timescale $\tau$ of the order of ten times the age of the universe. As $Q_1$ is so small, there is a large range of possible choices of $\tau_0$ so that  $\tau_0 \ll \tau < \tau_0/(Q_1 C_1)$ and so that SEC fails for times in $[0,\tau_0]$, from $10^{-68}\textrm{s}$
up to $10^{-2}\tau$, say.

Given these assumptions the approximation of Eq.~\eqref{eq:scen2approx} is reasonable, and our results show that a singularity is inevitable within about 10 times the age of the universe, for Hubble parameter
\be
H(0) > \frac{3B_1+1}{\tau} \sim 10^{-18} \textrm{s}^{-1} \,,
\ee
which on the order of the current values from PLANCK $(2.184 \pm 0.016) \times  10^{-18} \textrm{ s}^{-1}$~\cite{aghanim2018planck}. 

If the characteristic scale is based on the proton mass, $Q_0\tau^2\sim 0.1$ if $\tau$ is the age of the universe; however the minimum threshold on $H(0)$ is now $\sim 10^{-17} \textrm{s}^{-1}$, so the measured value would not be sufficient to conclude the existence of a singularity. For the Higgs field, $Q_0^{-1/2} \sim 2.36 \times 10^{6}$ years and the approximation $Q_0 \tau^2 \ll 1$ is only valid for $\tau$ less than $10^{-4}$ times the age of the universe, requiring therefore approximately $10^4$ times the measured Hubble parameter to infer that a singularity is inevitable in our toy model. 

These results give similar orders of magnitude for the required initial extrinsic curvature to those computed in Ref.~\cite{BrownFewsterKontou:2018} obtained with different methods (and phrasing the conditions for future timelike geodesic incompleteness). An advantage of the current method is that it can specify the timescale on which the focal points appear.

\subsection{The null case}

Now we turn to the null case. For any solution to the Einstein--Klein--Gordon equation in which the field magnitude obeys a global bound $|\phi|\leq \fmax < (8\pi \xi)^{1/2}$, it was shown in Ref.~\cite{Fewster:2010gm} that 
\be
\label{eqn:repineq}
\int_\gamma f^2\mathrm{Ric}(d\gamma,d\gamma) \leq 16\pi \xi \fmax^2 
\int_\gamma \left(\nabla_{d\gamma}\frac{f}{\sqrt{1-8\pi \xi \phi^2}} \right)^2  \,,
\ee
for all smooth compactly supported $(-\tfrac{1}{2})$-densities $f$ on $\gamma$,
where we have written the expression derived in
Ref.~\cite{Fewster:2010gm} invariantly and adapted it to our sign conventions.
Given any positive density $w$ on $\gamma$ this implies
\be
\label{eqn:ekgnull}
\int_\gamma f^2\mathrm{Ric}(d\gamma,d\gamma) \leq Q(\| \nabla_{d\gamma}f \|^2+\tilde{Q}(\gamma; w)^2 \| w f\|^2) \,,
\ee
where
\be\label{eq:QQtdefnull}
Q=\frac{32\pi \xi \fmax^2}{1-8\pi \xi \fmax^2} \,, \qquad \tilde{Q}(\gamma;w)=\frac{8\pi \xi \fmax }{1-8\pi \xi \fmax^2} \sup_\gamma \frac{|\nabla_{d\gamma} \phi|}{w}\,.
\ee 
Eq.~\eqref{eqn:ekgnull} is of the form of Eq.~\eqref{eq:Ricci_hypnull} with $m=1$, $Q_1(\gamma;w)=Q$ and $Q_0(\gamma;w)=Q\tilde{Q}(\gamma;w)^2$.  
Note that $Q_1$ in this case is independent of both $\gamma$ and $w$, while $Q_0(\gamma;w)$ is independent of any specific parametrisation of $\gamma$.

As described in Sec.~\ref{sub:focalnull} we fix the affine parametrization and $w$ in the following way: define $P$ to be a future converging spacelike submanifold of $M$ of co-dimension $2$ and let $\gamma$ be a future-directed null geodesic emanating normally from $P$. Extending $\hat{H}_\mu$ by parallel transport along $\gamma$ and choosing an affine coordinate $\lambda$, so that $p=\gamma(0)$ and $\hat{H}_\mu d\gamma^\mu/d\lambda =1$, we have $w=1$ and $q=\gamma(\ell)$, with $\ell=L_{\hat{H}}(\gamma)$ .

Then Eq.~\eqref{eqn:ekgnull} becomes
\be
\int_\gamma f^2(\lambda) R_{\mu \nu} U^\mu U^\nu d\lambda \leq Q(\gamma) \left( \| f' \|^2+\tilde{Q}^2(\gamma) \| f\|^2 \right) \,,
\ee
and
\begin{equation}
\tilde{Q}(\gamma)=\frac{8\pi \xi \fmax }{1-8\pi \xi \fmax^2} \sup_\gamma |\phi'(\lambda)| \,.
\end{equation}
Now we want to estimate $Q$ and $\tilde{Q}$. We consider a massless field and, as in the timelike case, we work in a hybrid model: a quantized scalar field in a thermal state of temperature $T$. In the massless scalar field case the Wick square of a KMS state with temperature $T$ is 
\be
\langle \nord{\phi^2} \rangle_T=\frac{T^{n-2}}{2^{n-2} \pi^{(n-1)/2}} \frac{\Gamma(n-2)}{\Gamma(\frac{n-2}{2})} \zeta(n-2) \,, 
\ee
where $\zeta$ is the Riemann zeta function. Similarly, if $U^\mu$ is any null
vector with $U^0=1$ then
\be
\langle \nord{(U^\mu \nabla_\mu \phi) (U^\nu \nabla_\nu \phi) } \rangle_T=\frac{T^n}{3 \cdot 2^{n-4} \pi^{(n-1)/2}} \frac{\Gamma(n)}{\Gamma(\frac{n-1}{2})} \zeta(n) \,.
\ee
For $\fmax^2 \sim \langle \nord{\phi^2} \rangle_T$, $\fmax'^2 \sim \langle \nord{(U^\mu \nabla_\mu \phi) (U^\nu \nabla_\nu \phi) } \rangle_T$ and restoring the units we have
\be
Q \sim (T/T_{\text{pl}})^{n-2} \,, \qquad \text{and} \qquad \tilde{Q} \sim Q \frac{kT}{\hbar} \,,
\ee
where $T_{\text{pl}}$ is the Planck temperature. 

Let us consider Scenario 1. For $n=4$, and a temperature $T \sim 10^{7} $K which is of the order of a newly formed neutron star~\cite{becker2009neutron} we have $Q_1=Q \sim 10^{-50} \ll 1$ and $Q_0 \sim 10^{-114} s^{-2}$. We can consider $\ell_0$ as a measurement in light seconds of
distance along null rays, measured by an observer at rest on the hypersurface $P$. Then assuming it is much larger than $10^{-57}$ light seconds, we have $Q_0 \ell_0^2 \ll 1$. 

The required magnitude of of the mean curvature vector of $P$ to form a focal point is given by Eq.~\eqref{eqn:meancurv} with the units restored
\be
- H \sim \frac{kT}{2 \hbar} \sqrt{8 A_1 B_1 Q^3} \sim 10^{-57} \textrm{s}^{-1} \,.
\ee
The contraction in Ref.~\cite{Fewster:2010gm} was found to be
\be
-H > \tilde{Q} \sqrt{Q(Q+(n-2))}+QK \coth{(K\ell_0)} \,,
\ee
where we have corrected some factors of $2$. Using the previous values of $Q$ and $\tilde{Q}$ and assuming that the second term which depends on the history of the solution does not get too large we get $-H > (kT/\hbar) Q^{3/2} \sqrt{2} \sim 10^{-57} \textrm{s}^{-1}$ which agrees with our estimate. 

The required mean normal curvature for this toy model is extremely small, scarcely more restrictive than being a trapped surface. Of course a model of a massless scalar should not be taken seriously as a model of astrophysical black hole formation which involves multiple species of interacting particles. However, it shows that a model where the NEC is be violated can still lead to geodesic incompleteness with very weak restrictions on the initial conditions. 

\section{Conclusion}
\label{sec:conclusions}

In this work we derived singularity theorems with weakened energy conditions inspired by QEIs, using index form methods. Compared to previous derivations that make use of the Raychaudhuri equation, our results provide simpler estimates of the required initial extrinsic curvature that leads to geodesic incompleteness. More importantly, in some cases, they give an estimate of the maximum proper time (in the timelike case), and affine parameter (in the null case) where the singularity is formed. 

The next step is to prove theorems with energy conditions derived directly from proven QEIs. In the timelike case the relevant QEI is the quantum strong energy inequality (QSEI) bounding the weighted renormalized effective energy density $T_{\mu\nu}U^\mu U_\nu-T/(n-2)$, the quantity appearing in the SEC. Such a QSEI was derived by the authors in a recent publication \cite{Fewster:2018pey} for the non-minimally coupled scalar field.

The null case presents greater challenges since no QEI along individual null rays is possible in four-dimensions \cite{Fewster:2002ne}. To overcome this, a promising approach is the technique of transverse smearing, averaging over a pencil of neighbouring null rays a few Planck lengths thick. Transverse smearing has been successfully used for the derivation of the averaged null energy condition (ANEC) \cite{Flanagan:1996gw, Kontou:2015yha}. However such versions of transversely smeared ANEC cannot be directly used in singularity theorems and new arguments are necessary.
 
If we are interested in utilising QEI bounds, we must consider that singularity theorems require bounds on the Ricci tensor rather than the stress-energy tensor. In the classical case, these are connected by the Einstein equation. In the quantum case and in the absence of a full theory of quantum gravity, a semiclassical approach could be employed. The semiclassical Einstein equation 
\be
\langle T_{\mu \nu} \rangle_\omega=-8\pi G_{\mu \nu} \,,
\ee
connects the expectation value of the renormalized stress-energy tensor with the classical Einstein tensor $G_{\mu \nu}$. This semiclassical approach to proving singularity theorems with hypotheses obeyed by quantum fields will be discussed elsewhere \cite{Fewster:unpublished}.

\begin{footnotesize}
	\noindent{\bf Acknowledgements}
We thank Atsushi Higuchi for useful comments on the text.
This work is part of a project that has received funding from the European Union's Horizon 2020 research and innovation programme under the Marie Sk\l odowska-Curie grant agreement No. 744037 ``QuEST''. 
\end{footnotesize} 

\appendix
\section{Calculations involving incomplete Beta functions} 
\label{appx:Sobolev}
In this appendix $\|\cdot\|$ will denote the standard $L^2$-norm on the unit interval $[0,1]$, except in~\eqref{eq:scalednorm}.

For $m\in\mathbb{N}$, we require values for the $L^2$-norms $\|p_m\|$, $\|p'_m\|$ and $\|p^{(m)}_m\|$ of the regularised incomplete Beta function
\begin{equation}
	p_m(t) := I(m,m;t) = \int_0^t g_{m-1}(s)\,ds ,
\end{equation}
where 
\begin{equation}\label{eq:gm}
g_m(t) = \mathcal{N}_m t^m (1-t)^m , \qquad \mathcal{N}_m = B(m+1,m+1)^{-1} = \frac{(2m+1)!}{m!^2}.
\end{equation}
Noting that $g_m^{(k)}(0)=g^{(k)}_m(1)=0$ for $0\le k\le m-1$, $p_m$ is a non-decreasing polynomial of degree $2m-1$ obeying
\[
p_m(0)=0,\qquad p_m(1)=1,\qquad p_m^{(k)}(0)=p^{(k)}_m(1)=0 \quad (1\le k\le m-1).
\]

Starting with $\|p_m\|^2$, direct calculation gives the values  $A_1=1/3$, $A_2=13/35$, $A_3=181/462$, $A_4=521/1287$ stated in the text. The general closed form expression,  
\begin{equation}\label{eq:fmnorm2}
\|p_m\|^2=A_m:= \frac{1}{2} - \frac{(2m)!^4}{4(4m)!m!^4},
\end{equation}
will be derived elsewhere, along with various other exact formulae for integrals of products of incomplete beta functions~\cite{ConnorFewster:2019}. It follows from~\eqref{eq:fmnorm2} that $\|p_m\|^2\in [1/3,1/2)$, with $\|p_m\|^2\to 1/2$ as $m\to\infty$. 

Next, as $p'_m =g_{m-1}$, we find immediately that
\begin{equation}
\|p'_m\|^2 = B_m:= \frac{B(2m-1,2m-1)}{B(m,m)^2}\sim \sqrt{\frac{2m}{\pi}}
\end{equation}
as $m\to\infty$. We record the values $B_1=1$, $B_2=6/5$, $B_3=10/7$, $B_4=700/429$. 

It is not clear whether $\| p_m^{(k)} \|^2$ can in general be expressed in a simple closed form but perhaps surprisingly, $\|p_m^{(m)}\|^2$ can. To do this, note first that $\|p_m^{(m)}\|^2=\|g_{m-1}^{(m-1)}\|^2$. Using the boundary conditions noted above, we can integrate by parts $m$ times to find
\[
\|g_m^{(m)}\|^2= (-1)^m \int_0^1 g_m^{(2m)}(t) g_m(t)\,dt = \mathcal{N}_m(2m)! = \frac{(2m)!(2m+1)!}{m!^2}\sim e^{-1}(4m/e)^{2m-1}. 
\]
Here we have used the fact that $g^{(2m)}(t)=(-1)^m(2m)!$ as is clear from the definition~\eqref{eq:gm}. Thus
\begin{equation}\label{eq:AmBm} 
\|p_m^{(m)}\|^2= C_m: =  \frac{(2m-2)!(2m-1)!}{(m-1)!^2} .
\end{equation}
We record the values $C_1=1$, $C_2=12$, $C_3=720$, $C_4=100800$.

Now consider an interval $[0,\tau]$. Now writing $P_m(t)=f_{m}(t/\tau)$, we have
$P_m^{(k)}(0)=0$ ($0\le k\le m-1$), $P_m(\tau)=1$, $P^{(k)}(\tau)=0$ ($1\le k\le m-1$),
and clearly 
\begin{equation}\label{eq:scalednorm}
\|P_m^{(k)}\|^2 = \frac{\|p_m^{(k)}\|^2}{\tau^{2k-1}}, 
\end{equation}
where the norm on the left-hand side is now taken on $[0,\tau]$.

\providecommand{\apj}{Astrophys. J.\ }
\providecommand{\nat}{Nature (London)\ }
\end{document}